\documentclass[sigconf]{acmart}

\AtBeginDocument{%
  \providecommand\BibTeX{{%
    \normalfont B\kern-0.5em{\scshape i\kern-0.25em b}\kern-0.8em\TeX}}}

\copyrightyear{2024}
\acmYear{2024}

\setcopyright{acmlicensed}\acmConference[WWW '24 Companion]{Companion Proceedings of the ACM Web Conference 2024}{May 13--17, 2024}{Singapore, Singapore}
\acmBooktitle{Companion Proceedings of the ACM Web Conference 2024 (WWW '24 Companion), May 13--17, 2024, Singapore, Singapore}

\acmISBN{979-8-4007-0172-6/24/05}
\acmDOI{10.1145/3589335.3648327}

\usepackage{amsmath}
\usepackage{graphicx}
\usepackage{multirow}
\usepackage{subfigure}
\usepackage{float}
\usepackage{bbding}
\usepackage{pifont}
\definecolor{midnightgreen}{rgb}{0.0, 0.29, 0.33}

\newcommand{\sysname}{{MS MARCO Web Search}}

\begin{document}

\title{\sysname{}: a Large-scale Information-rich Web Dataset with Millions of Real Click Labels}

\author{Qi Chen}
\affiliation{
\institution{Microsoft}
\country{Beijing, China}
}
\author{Xiubo Geng}
\affiliation{
\institution{Microsoft}
\country{Beijing, China}
}
\author{Corby Rosset}
\affiliation{
\institution{Microsoft}
\country{Redmond, United States}
}
\author{Carolyn Buractaon}
\affiliation{
\institution{Microsoft}
\country{Redmond, United States}
}
\author{Jingwen Lu}
\affiliation{
\institution{Microsoft}
\country{Redmond, United States}
}
\author{Tao Shen}
\authornote{The work was done at Microsoft.}
\affiliation{
\institution{University of Technology Sydney}
\country{Sydney, Australia}
}
\author{Kun Zhou}
\affiliation{
\institution{Microsoft}
\country{Beijing, China}
}
\author{Chenyan Xiong$^{\ast}$}
\affiliation{
\institution{Carnegie Mellon University}
\country{Pittsburgh, United States}
}
\author{Yeyun Gong}
\affiliation{
\institution{Microsoft}
\country{Beijing, China}
}
\author{Paul Bennett$^{\ast}$}
\affiliation{
\institution{Spotify}
\country{New York, United States}
}
\author{Nick Craswell}
\affiliation{
\institution{Microsoft}
\city{Redmond}
\country{United States}
}
\author{Xing Xie}
\affiliation{
\institution{Microsoft}
\city{Beijing}
\country{China}
}
\author{Fan Yang}
\affiliation{
\institution{Microsoft}
\city{Beijing}
\country{China}
}
\author{Bryan Tower}
\affiliation{
\institution{Microsoft}
\city{Redmond}
\country{United States}
}
\author{Nikhil Rao}
\affiliation{
\institution{Microsoft}
\city{Mountain View}
\country{United States}
}
\author{Anlei Dong}
\authornote{Author names listed alphabetically by surname.}
\affiliation{
\institution{Microsoft}
\city{Mountain View}
\country{United States}
}
\author{Wenqi Jiang$^{\dagger}$}
\affiliation{
\institution{ETH Zürich}
\city{Zürich}
\country{Switzerland}
}
\author{Zheng Liu$^{\dagger}$}
\affiliation{
\institution{Microsoft}
\city{Beijing}
\country{China}
}
\author{Mingqin Li$^{\dagger}$}
\affiliation{
\institution{Microsoft}
\city{Redmond}
\country{United States}
}
\author{Chuanjie Liu$^{\dagger}$}
\affiliation{
\institution{Microsoft}
\city{Beijing}
\country{China}
}
\author{Zengzhong Li$^{\dagger}$}
\affiliation{
\institution{Microsoft}
\city{Redmond}
\country{United States}
}
\author{Rangan Majumder$^{\dagger}$}
\affiliation{
\institution{Microsoft}
\city{Redmond}
\country{United States}
}
\author{Jennifer Neville$^{\dagger}$}
\affiliation{
\institution{Microsoft}
\city{Redmond}
\country{United States}
}
\author{Andy Oakley$^{\dagger}$}
\affiliation{
\institution{Microsoft}
\city{Redmond}
\country{United States}
}
\author{Knut Magne Risvik$^{\dagger}$}
\affiliation{
\institution{Microsoft}
\city{Oslo}
\country{Norway}
}
\author{Harsha Vardhan Simhadri$^{\dagger}$}
\affiliation{
\institution{Microsoft}
\city{Bengaluru}
\country{India}
}
\author{Manik Varma$^{\dagger}$}
\affiliation{
\institution{Microsoft}
\city{Bengaluru}
\country{India}
}
\author{Yujing Wang$^{\dagger}$}
\affiliation{
\institution{Microsoft}
\city{Beijing}
\country{China}
}
\author{Linjun Yang$^{\dagger}$}
\affiliation{
\institution{Microsoft}
\city{Redmond}
\country{United States}
}
\author{Mao Yang$^{\dagger}$}
\affiliation{
\institution{Microsoft}
\city{Beijing}
\country{China}
}
\author{Ce Zhang$^{\ast}$$^{\dagger}$}
\affiliation{
\institution{ETH Zürich}
\city{Zürich}
\country{Switzerland}
}

\renewcommand{\shortauthors}{Qi Chen et al.}

\begin{abstract}
Recent breakthroughs in large models have highlighted the critical significance of data scale, labels and modals. In this paper, we introduce \sysname{}, the first large-scale information-rich web dataset, featuring millions of real clicked query-document labels. This dataset closely mimics real-world web document and query distribution, provides rich information for various kinds of downstream tasks and encourages research in various areas, such as generic end-to-end neural indexer models, generic embedding models, and next generation information access system with large language models. \sysname{} offers a retrieval benchmark with three web retrieval challenge tasks that demands innovations in both machine learning and information retrieval system research domains. As the first dataset that meets large, real and rich data requirements, \sysname{} paves the way for future advancements in AI and system research. MS MARCO Web Search dataset is available at: {\color{blue}{\url{https://github.com/microsoft/MS-MARCO-Web-Search}}}.
\end{abstract}



\maketitle

\section{Introduction}

Recently, the large language model (LLM), a breakthrough in the field of artificial intelligence, has provided a novel way for people to access information through interactive communication. Although it has become an indispensable tool for tasks such as content creation, semantic understanding and conversational AI, it still exhibits certain limitations. One such limitation is the model's tendency to produce hallucinated or fabricated content, as it generates responses based on patterns observed in the training data rather than verifying factual accuracy. Furthermore, it struggles with real-time knowledge updates, as it can only provide information available up until the time of its last training. This makes it less reliable for retrieving the latest, dynamic information. Therefore, integrating an external up-to-date knowledge base with large language models is of paramount importance to enhance their performance and reliability. This combination not only mitigates the limitations of hallucination and knowledge update but also broadens the model's applicability across various domains, making it more versatile and valuable. Consequently, information retrieval systems, like the Bing search engine~\cite{bing}, continue to play a vital role in the new LLM-based information systems, such as Webgpt~\cite{nakano2021webgpt} and new Bing~\cite{newbing}.

For modern information retrieval systems, the core is the large semantic understanding model, such as a neural indexer model~\cite{wang2022neural} or dual embedding model~\cite{huang2013learning,shen2014learning,palangi2016deep,hu2014convolutional,devlin2018bert,qiao2019understanding,reimers2019sentence,shan2021glow,xiong2020approximate}, which can capture users' intents as well as the rich meanings of a document with better tolerance for out of vocabulary words, spelling errors, and synonymous expressions. Training a high-quality large semantic understanding model requires a vast amount of data to achieve sufficient knowledge coverage. The larger the dataset, the better the model is likely to perform, as the model can learn more complex and sophisticated patterns and correlations.

High-quality human-labeled data is as important as data scale. Recent research, such as InstructGPT~\cite{ouyang2022training} and LLAMA-2~\cite{touvron2023llama}, has demonstrated the crucial role of labeled data for training large foundation models. These models rely on large volumes of training data to learn generalizable features, while human-labeled data enable the model to learn the specific tasks it is designed for. This also applies to large semantic understanding models.

Moreover, information-rich data is also crucial for training large semantic understanding models effectively. The use of multi-modal datasets can help models understand complex relationships between different types of data and transfer knowledge between them. For example, using images and text in a multi-modal data set can help models learn about image concepts and their corresponding text descriptions, providing a more holistic representation of the data. 

\begin{table*}[t]
  \centering
  \caption{Comparison of \sysname{} (with ClueWeb22) and existing datasets}
  \begin{tabular}{l|rrcccc}
    \hline 
    Dataset & \#Documents & \#Queries & Web & Multi-lingual docs & Rich-info docs & Multi-lingual queries \\
    \hline
    Robust04 & 528K & 250 & -  & - & - & - \\
    ClueWeb09 & 1B & - & \Checkmark & \Checkmark & - & -\\
    ClueWeb12 & 733M & - & \Checkmark & - & - & - \\
    GOV2 & 25M & 50 & - & - & - & - \\
    Common Crawl (One Dump) & 3.1B & - & \Checkmark & \Checkmark & - & -\\
    Natural Questions & 28M & 320K & - & - & - & -\\
    MS MARCO & 3.2M & 100K & - & - & - & - \\
    MS MARCO Ranking v2 & 11M & 1M & - & - & - & - \\ 
    ORCAS & 3.2M & 10M & - & - & - & - \\
    CLIR & 23.9M & 2.8M & - & \Checkmark & - & - \\
    \hline 
    \sysname{} (w. ClueWeb22) & \bf{10B} & \bf{10M} & \Checkmark & \Checkmark & \Checkmark & \Checkmark \\
    \hline 
  \end{tabular}
  \label{tab:datasets}
\end{table*}

The emerging large, real and rich data requirements motivate us to create a new \sysname{} dataset, the first large-scale information-rich web dataset with millions of real clicked query-document labels. \sysname{} incorporates the largest open web document dataset, ClueWeb22~\cite{overwijk2022clueweb22}, as our document corpus. ClueWeb22 includes about 10 billion high-quality web pages, sufficiently large to serve as representative web-scale data. It also contains rich information from the web pages, such as visual representation rendered by web browsers, raw HTML structure, clean text, semantic annotations, language and topic tags labeled by industry document understanding systems, etc. \sysname{} further contains 10 million unique queries from 93 languages with millions of relevant labeled query-document pairs collected from the search log of the Microsoft Bing search engine to serve as the query set. This large collection of multi-lingual information-rich real web documents, queries and labeled query-document pairs enables various kinds of downstream tasks and encourages several new research directions that previous datasets cannot well support, for example, generic end-to-end neural indexer models, generic embedding models, and next generation information access system with large language models, etc. As the first large, real and rich web dataset, \sysname{} will serve as a critical data foundation for future AI and systems research. 

\sysname{} offers a retrieval benchmark which implements several state-of-the-art embedding models, retrieval algorithms, and retrieval systems originally developed on existing datasets. We compare the quality of their results and system performance on our new \sysname{} dataset as the benchmark baselines for web scale information retrieval. The experiment results demonstrate that embedding models, retrieval algorithms, and retrieval systems are all critical components in web information retrieval. And interestingly, simply improving only one component may bring negative impacts to the end-to-end retrieval result quality and system performance. We hope that this retrieval benchmark can facilitate future innovations in data-centric techniques, embedding models, retrieval algorithms, and retrieval systems to maximize end-to-end performance. 

\section{Background and Related Work}

\subsection{Web Scale Information Retrieval}
In traditional information retrieval, user queries and documents are represented as a list of keywords, and the retrieval is done based on keyword matching. However, simple keyword matching faces many challenges. First, it cannot clearly understand users' intents. In particular, it cannot estimate users' positive and negative sentiment and may return opposite results by mistake. Second, it cannot combine synonymous expressions, reducing the diversity of results~\cite{guo2022semantic}. Third, it cannot handle spelling errors and will return irrelevant results. Therefore, query alteration is employed to address the above challenges. Unfortunately, it is difficult to cover all kinds of query alterations, especially those newly-appeared alterations.

With the great success of deep learning in natural language processing, both queries and documents can be more meaningfully represented as semantic embedding vectors. Since embedding-based retrieval solves the above three challenges, it has been widely used in modern information systems to facilitate new state-of-the-art retrieval quality and performance. Numerous prior studies have concentrated on deep embedding models, from DSSM~\cite{huang2013learning}, CDSSM~\cite{shen2014learning}, LSTM-RNN~\cite{palangi2016deep}, and ARC-I~\cite{hu2014convolutional} to transformer-based embedding models~\cite{devlin2018bert,reimers2019sentence,qiao2019understanding,shan2021glow,xiong2020approximate,chen2024bge,xiao2023c}. They have shown impressive gains with brute-force nearest neighbor embedding search on some small datasets as compared with traditional keyword matching.

Due to the extremely high computational cost and query latency of brute-force vector search, there are many research approaches focusing on large-scale approximate vector nearest neighbor search (ANN) algorithms and systems design~\cite{jegou2011searching,faiss17,baranchuk2018revisiting,babenko2014inverted,lopq2014locally,lopq2014locally,gnoimi2016efficient, jayaram2019rand,scann2020,hmann2020,chen2021spann}. They can be divided into partition-based and graph-based solutions. Partition-based solutions, such as SPANN~\cite{chen2021spann}, divide the whole vector space into a large number of clusters and only do fine-grained search on a small number of the closest clusters to a query in online search. Graph-based solutions, such as DiskANN~\cite{jayaram2019rand}, construct a neighbor graph for the whole dataset and do the best-first traversal from some fixed starting points when a query comes in. Both of these approaches work well on some uniform-distributed datasets.

Unfortunately, when applying embedding-based retrieval in the web scenario, several new challenges emerge. First, web scale data volumes require large models, high embedding dimensions, and a large-scale labeled training dataset to guarantee sufficient knowledge coverage. 
Second, performance gains of state-of-the-art embedding models verified on small datasets cannot be directly transferred to a web scale dataset (see section~\ref{sec:embedding_model_eval}).
Third, embedding models need to co-work with ANN systems in order to serve large scale data volumes efficiently. However, different training data distributions may affect the accuracy and system performance of an ANN algorithm, which will greatly reduce the result accuracy as compared to embedding models with brute-force search. Distill-VQ~\cite{xiao2022distill} has verified that CoCondenser~\cite{gao2022unsupervised} embedding model with Faiss-IVFPQ ANN index achieves different result accuracy on MSMarco~\cite{msmarco} and NQ~\cite{nq} datasets. Moreover, even the same training data distribution will also result in different embedding vector distributions, which will lead to different ranking trends of the embedding models in brute-force search (KNN) and approximate nearest neighbor search (ANN) (see section~\ref{sec:end-to-end}).

\subsection{Existing Datasets}
To encourage innovation in the information retrieval area, the community has collected several datasets for public benchmarking (summarized in Table~\ref{tab:datasets}).

There are many public web datasets for traditional information retrieval tasks, such as Robust04~\cite{Robust04}, ClueWeb09~\cite{clarke2009overview}, ClueWeb12~\cite{callan2012lemur}, GOV2~\cite{Clarke2004TrecTerabyte}, ClueWeb22~\cite{overwijk2022clueweb22} and Common Crawl~\cite{commoncrawl}.  Unfortunately, these datasets have, at most, hundreds of labeled queries, far from enough to learn a good deep learning enhanced retrieval model. 

Recently, several new datasets have been published for research on deep learning enhanced retrieval~\cite{msmarco,nq,sasaki2018cross}. MS MARCO~\cite{msmarco} is one of the most popular datasets for embedding model investigation. It provides 100K questions collected from Bing’s search questions paired with human generated answers contextualized within web documents. MS MARCO Ranking v2~\cite{soboroff2021overview} expands the size of the document and question sets to 11 million and 1 million, respectively. ORCAS~\cite{craswell2020orcas} provides 10 million unique queries and 18 million clicked query-document pairs for MS MARCO documents. Natural Questions~\cite{nq}, a sub-million-scale question answering dataset collected from Google’s search queries with human annotated answers in Wikipedia articles, was repurposed for embedding-based retrieval by extracting passages from Wikipedia as candidate answers~\cite{karpukhin2020dense}. CLIR~\cite{sasaki2018cross}, a million-scale cross-language information retrieval dataset collected from Wikipedia, has been used to train cross-lingual embedding models~\cite{xu2021leveraging}. However, none of these datasets meet the emerging large, real and rich requirements. These datasets focus on English-only question answering tasks. None of them has the desired web-scale data with highly-skewed multi-lingual queries which can be short, ambiguous and often not formulated as natural language questions. Further, they only provide the raw text of queries and answers, which limits the potential of future cross-modal knowledge transfer research. Finally, they only focus on evaluating the quality of embedding models using brute-force search, which cannot reflect end-to-end retrieval challenges.

ANN benchmark~\cite{aumuller2017ann} and Billion-scale ANN benchmark ~\cite{bigann} provide multiple high-dimensional vector datasets to evaluate the result accuracy and system performance for embedding-based retrieval algorithms. Unfortunately, they cannot measure model quality and thus cannot reflect the end-to-end retrieval performance.

Therefore, a large-scale information-rich web dataset with real document and query distribution that can reflect real-world challenges is still lacking.

\begin{figure}
\centering
\includegraphics[width=1.0\linewidth]{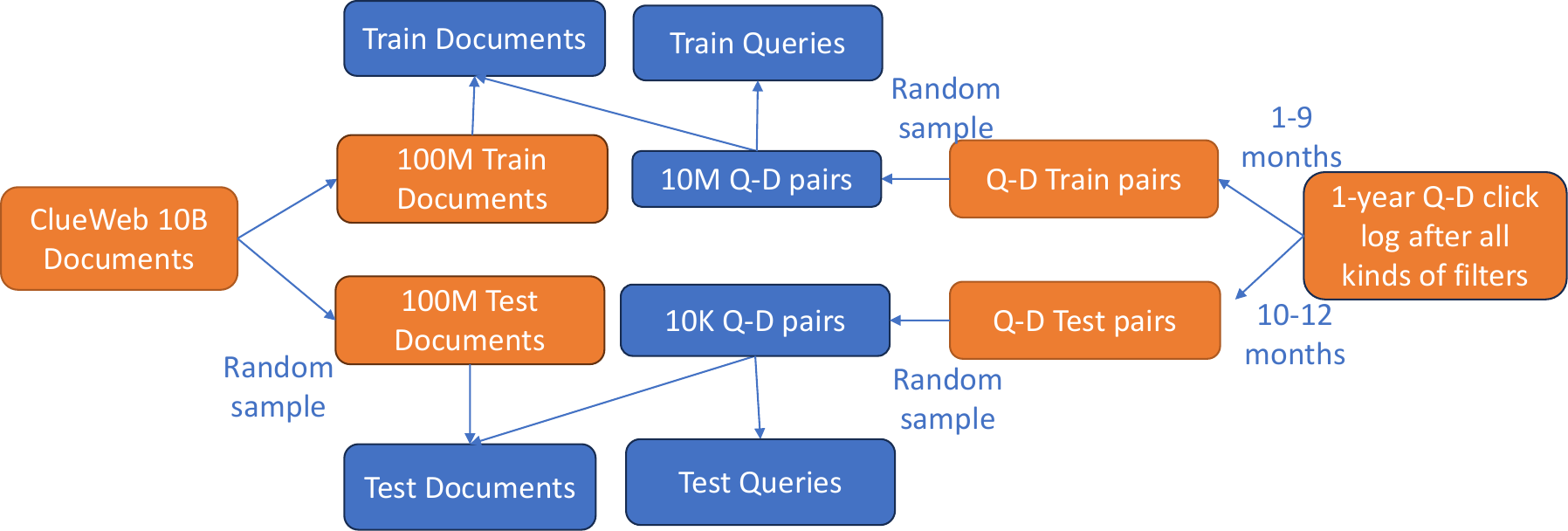}
\caption{The creation of the \sysname{} dataset}
\label{fig:data_creation}
\end{figure}

\begin{figure*}
\centering
\includegraphics[width=0.8\linewidth]{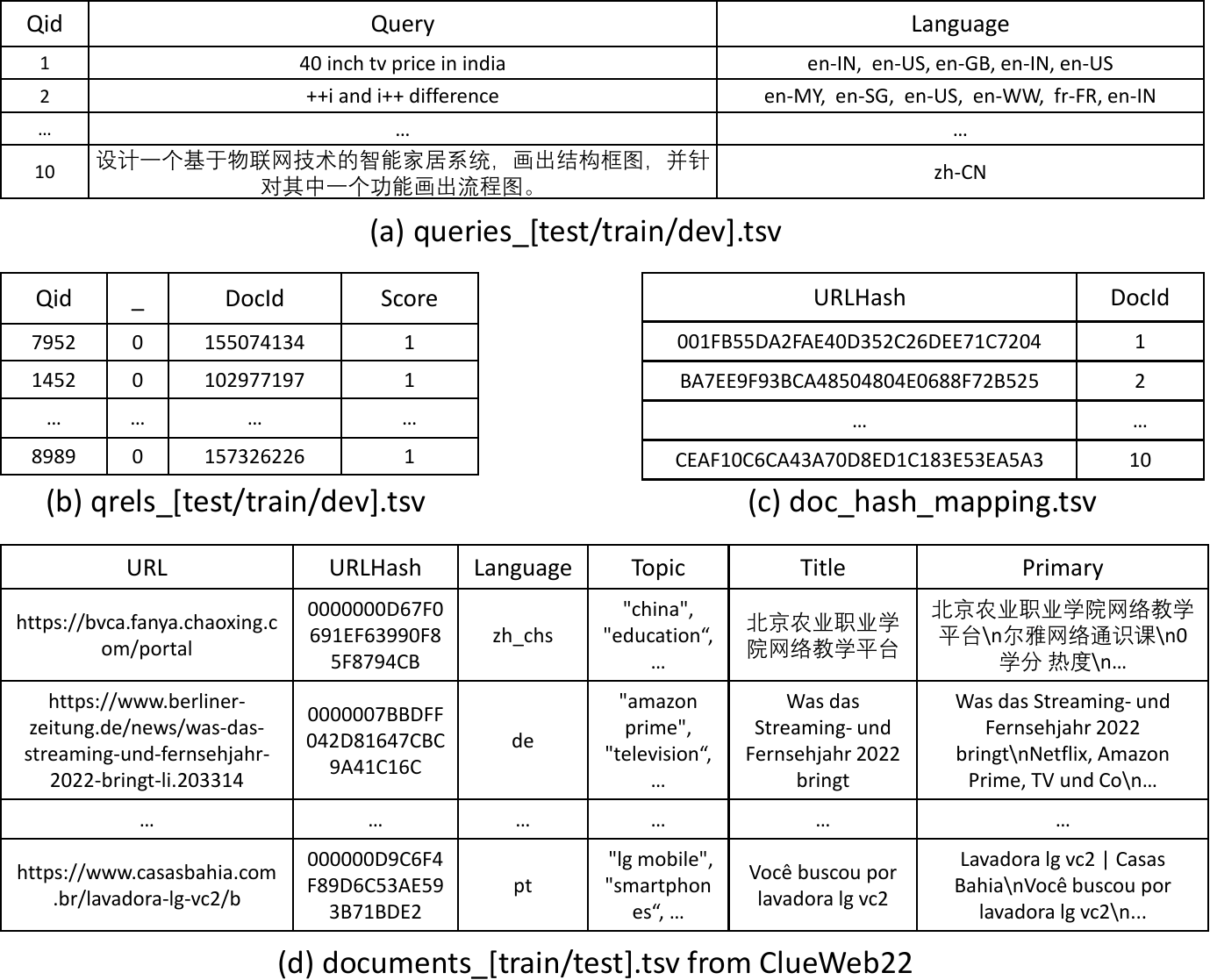}
\caption{The example files in the \sysname{} dataset
}
\label{fig:data_example}
\end{figure*}

\section{\sysname{} Dataset}
In this paper, we present \sysname{}, a large-scale dataset for research on web information retrieval. \sysname{} dataset consists of a high quality set of web pages that mirrors the highly-skewed web document distribution, a query set that reflects the real web query distribution, and a large-scale query-document label set for embedding model training and evaluation.

\subsection{Document Preparation}
We use ClueWeb22~\cite{callan2012lemur} as our document set since it is the largest and newest open web document dataset for our purpose. It meets the requirements of large scale, high quality and realistic document distributions crawled and processed by a commercial web search engine with rich information. Compare to Common Crawl~\cite{commoncrawl} which only crawls 35 million registered domains and covers 40+ languages, ClueWeb22 closely mimics the realistic crawl selection of a commercial search engine with 207 languages. It has 10 billion high-quality web pages with rich affiliated information, such as url, language tag, topic tag, title and clean text, etc. Figure ~\ref{fig:data_example} (d) gives an example of the data structures provided by ClueWeb22.

To make training cost-effective for both academia and industry, we provide a 100 million and a 10 billion document set. The 100 million document set is a random subset of the 10 billion document set. In order to evaluate model generalization ability in a small-scale dataset, two 100 million non-overlapping document sets are provided, one for training and the other for testing. The whole process is shown in the left part of figure \ref{fig:data_creation}.

\subsection{Query Selection and Labeling}
To generate large scale high quality queries and query-document relevance labels, we sample query-document clicks from one year of Bing search engine's logs. The initial query set gets filtered to remove queries that are rarely triggered, contain personally identifiable information, offensive content, adult content and those having no click connection to the ClueWeb22 document set. The resulting set includes queries triggered by many users, which reflects the real query distribution of a commercial web search engine. 

The queries are split into train and test sets based on time, which is similar to real-world web scenarios training an embedding model using past data and serving future incoming web pages and queries. We sample around 10 million query-document pairs from the train set and 10 thousand query-document pairs from the test set. The documents in the query-document train and test sets are then merged into the 100 million train document set and test document set respectively (shown in right part of figure \ref{fig:data_creation}). To enable quality verification of the model during training, we split a dev query-document set from the train query-document set. Since the train and dev sets share the same document set, the dev set can be used to quickly verify the training correctness and model quality during training. For the 10B dataset, we use the same train, dev, and test queries but sample more query-document pairs. 

\subsection{Dataset Analysis}

\begin{table*}[]
  \centering
  \caption{100M and 10B dataset statistics}
  \begin{tabular}{|l|c|c|c|c|c|c|c|c|c|}
    \hline
    \multirow{2}{*}{Scale} & \multicolumn{3}{c|}{Train} & \multicolumn{3}{c|}{Dev} & \multicolumn{3}{c|}{Test} \\ \cline{2-10}
                         ~ & Document & Query & Q-D & Document & Query & Q-D & Document & Query & Q-D \\ \hline
    \hline
    Set-100M &  109969872 & 9206475 & 9346695 & 109969872 & 9253 & 9402 & 100924960 & 9374 & 9374 \\
    Set-10B &  10B & 9206475 & 62302553  &  10B & 9253 & 63314 &  10B & 9374 & 40511 \\
    \hline
  \end{tabular}
  \label{tab:dataset_overview}
\end{table*}

We have constructed two scales of the datasets: Set-100M and Set-10B. Table \ref{tab:dataset_overview} gives the detailed statistics of the datasets. The example files of \sysname{} Set-100M are shown in figure \ref{fig:data_example}.

\subsubsection{Language Distribution Analysis}
\label{sec:lang_analysis}
\begin{figure*}
\centering
\subfigure[Train]{
    \begin{minipage}[b]{.30\textwidth}
    \label{fig:query_lang_train}
    \includegraphics[width=1\linewidth]{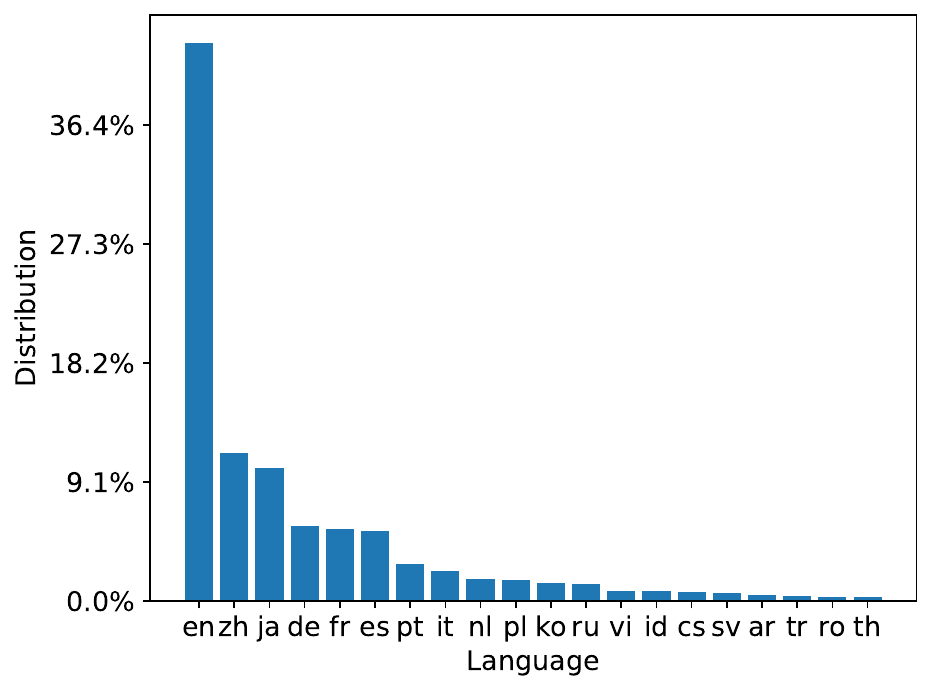}
    \end{minipage}
}
\subfigure[Test]{
    \begin{minipage}[b]{.30\textwidth}
      \label{fig:query_lang_test}
    \includegraphics[width=1\linewidth]{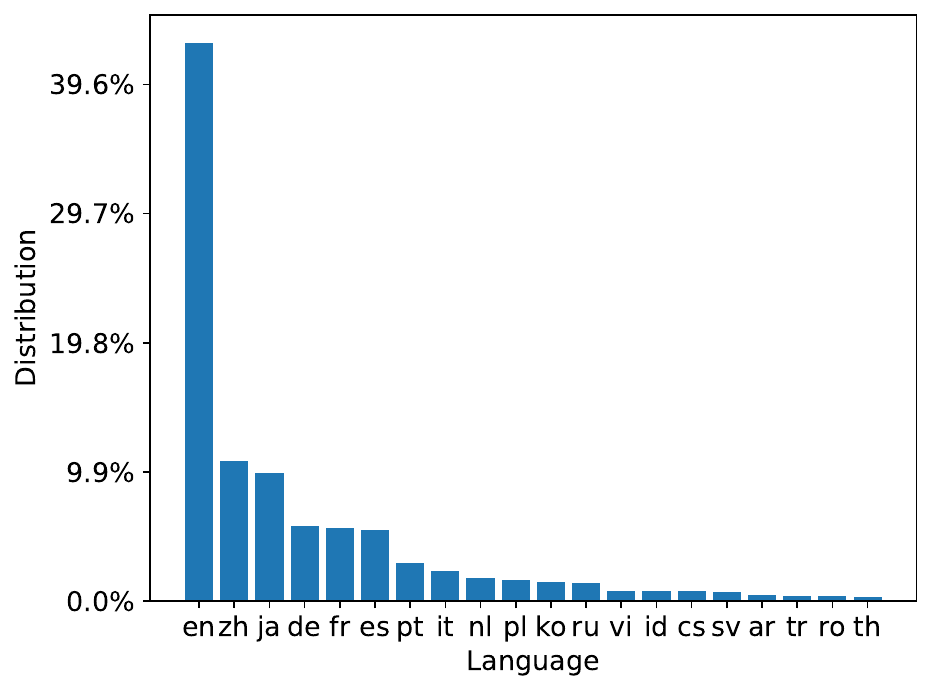}
    \end{minipage}
}
\caption{
The distribution of 20 most popular languages in the train and test document sets.
}
\label{fig:corpus_lang}
\end{figure*}

\begin{figure*}
\centering
\subfigure[Train]{
    \begin{minipage}[b]{.30\textwidth}
    \label{fig:query_lang_train}
    \includegraphics[width=1\linewidth]{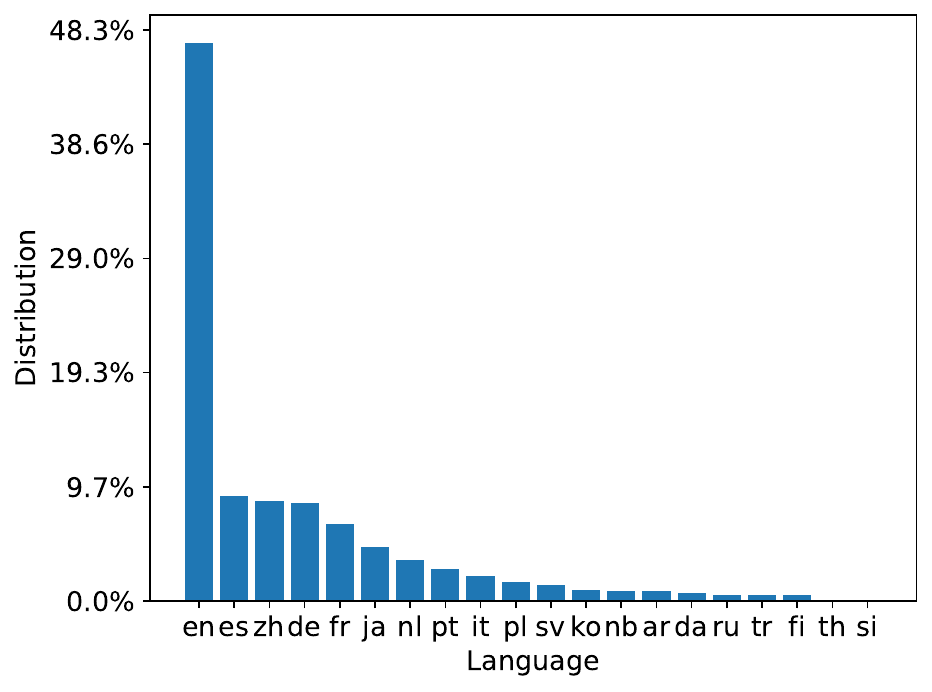}
    \end{minipage}
}
\subfigure[Dev]{
    \begin{minipage}[b]{.30\textwidth}
    \label{fig:query_lang_dev}
     \includegraphics[width=1\linewidth]{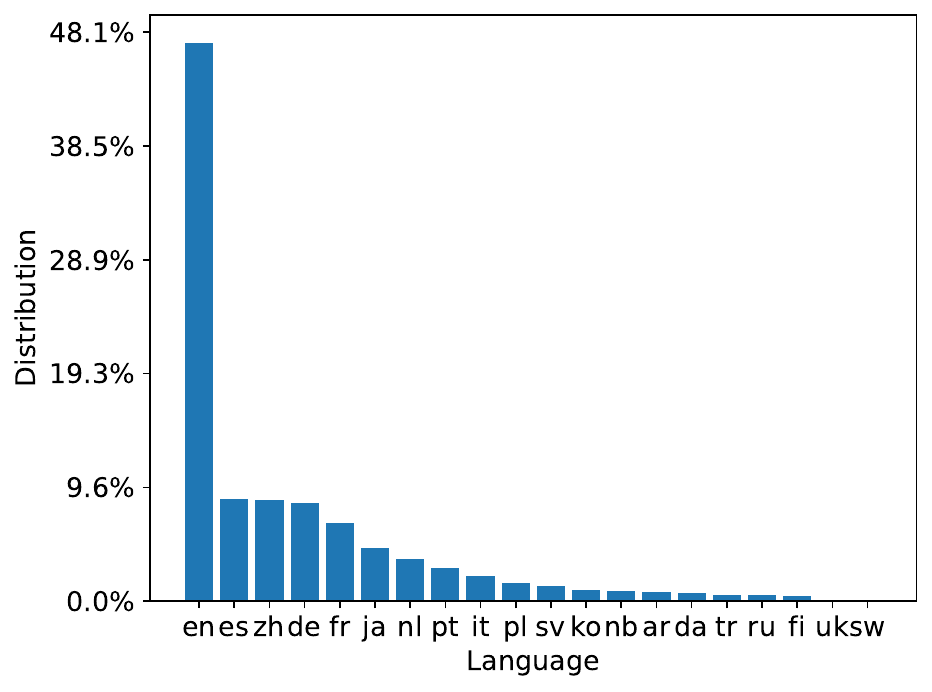}
     \end{minipage}
}
\subfigure[Test]{
    \begin{minipage}[b]{.30\textwidth}
      \label{fig:query_lang_test}
    \includegraphics[width=1\linewidth]{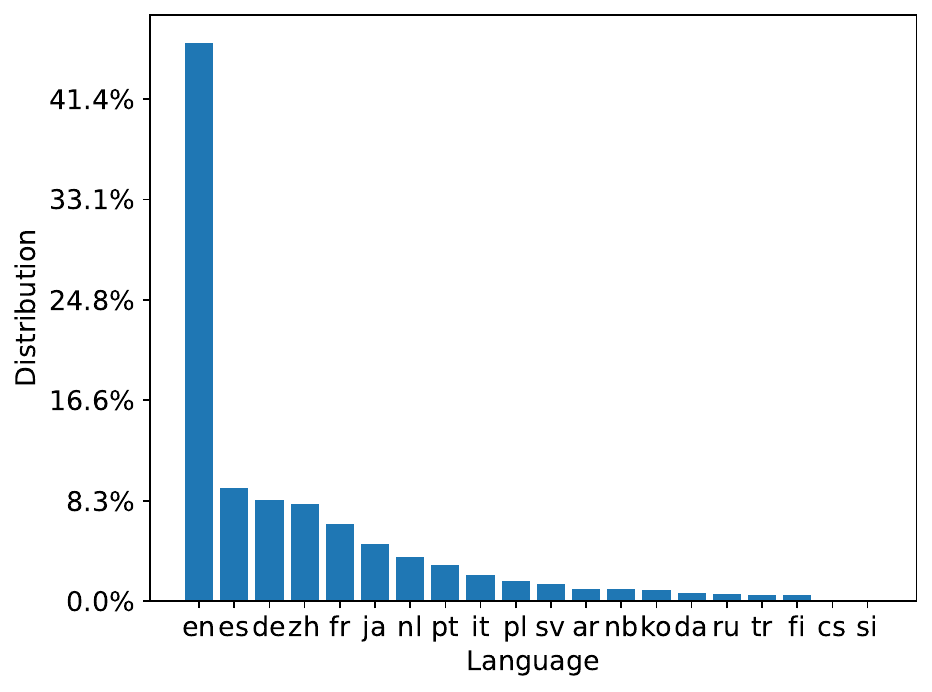}
    \end{minipage}
}
\caption{
The distribution of 20 most popular languages in the train, dev and test query sets.
}
\label{fig:query_lang}
\end{figure*}

\sysname{} is a multi-lingual dataset with its queries and document both from a commercial web search engine. We analyze the 20 most popular languages among 93 and 207 languages in both queries and documents in the 100M dataset respectively; the 10B dataset has a similar distribution. Figure \ref{fig:corpus_lang} summarizes the document language distribution in the train and test document sets. We can see that both train and test document sets are aligned with the original ClueWeb22 document distribution. Figure \ref{fig:query_lang} summarizes the query language distribution in the train, dev, and test query sets. From the distribution, we can see that the language distribution of the queries in the web scenario is high-skewed which may lead to model bias. It encourages research on data-centric techniques for training data optimization.

\subsubsection{Data Skew Analysis} 
\begin{figure*}
\centering
\subfigure[]{
    \begin{minipage}[b]{.31\textwidth}
    \label{fig:query_corpus}
    \includegraphics[width=1\linewidth]{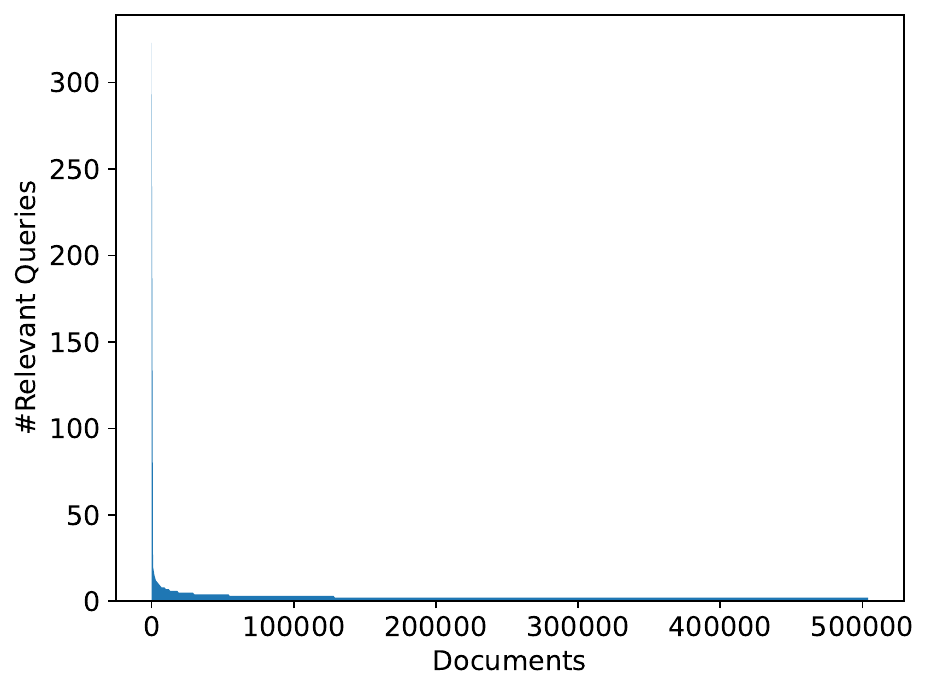}
    \end{minipage}
}
\subfigure[]{
    \begin{minipage}[b]{.31\textwidth}
      \label{fig:corpus_query}
    \includegraphics[width=1\linewidth]{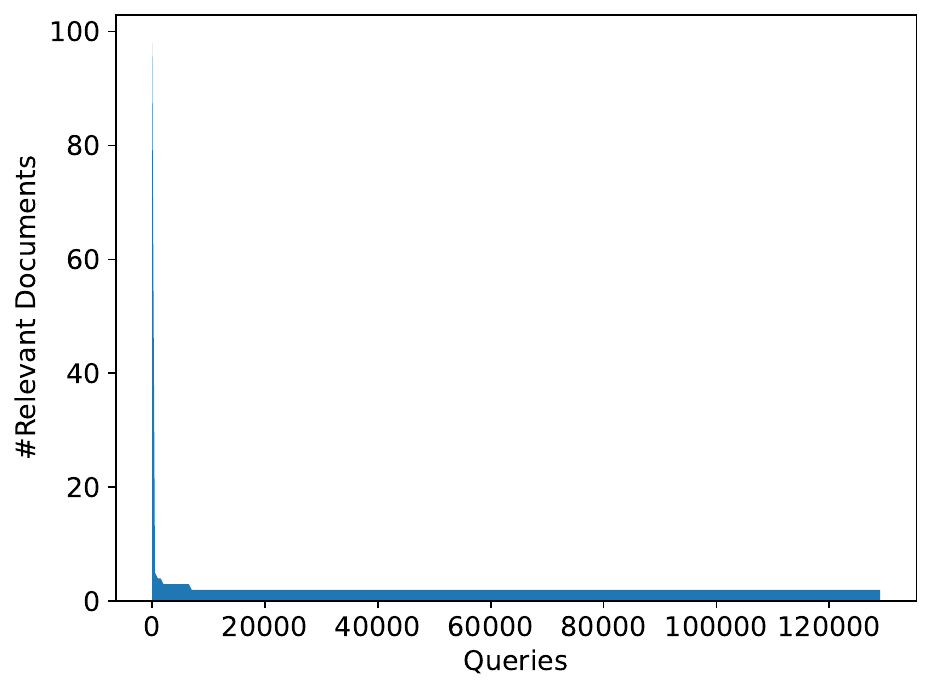}
    \end{minipage}
}
\caption{
The distribution of documents and queries labels
}
\label{fig:q_c_d}
\end{figure*}

We analyze the query-document label distribution in the training data. Figure \ref{fig:query_corpus} shows documents and the number of relevant queries associated with them. From the figure, we can see that there are only a few documents with multiple labels: only 7.77\% of the documents have relevant labeled queries and 0.46\% of documents have more than one labeled relevant query. Figure \ref{fig:corpus_query} summarizes the queries and their relevant documents. From the figure, we can see that only 1.4\% of queries have multiple relevant documents. This highly skewed nature of the dataset is consistent with what is observed while training models for web-scale information retrieval. Our intention is to keep this skew to make models trained on this dataset applicable to real-world scenarios.

\subsubsection{Test-Train Overlap Analysis}

\begin{table*}[!htbp]
  \centering
  \caption{Test query-document pair distribution}
  \begin{tabular}{|l|c|c|c|c|}
    \hline
    Total & Q$\in$Train, C$\in$Train & Q$\notin$Train, C$\in$Train & Q$\in$Train, C$\notin$Train & Q$\notin$Train, C$\notin$Train  \\
    \hline
    9374 &  82  & 1468 & 178 & 7646 \\
    \hline
  \end{tabular}
  \label{tab:qc_distribution}
\end{table*}

As introduced in~\cite{lewis2021question}, there exists large test-train overlap in some popular
open-domain QA datasets, which cause many popular open-domain models to simply memorize the queries seen at the training stage. Subsequently, they perform worse on novel queries. The work~\cite{zhan2022evaluating} observes this phenomenon in the MSMARCO dataset. To better evaluate  model generalizability, we minimize the overlap between the train and test sets by splitting the query-document pairs into train and test sets by time. This means the test query-document pairs have no time overlap with the train query-document pairs, which introduces a large portion of novel queries. This can be verified in the table \ref{tab:qc_distribution}. We summarize the test query-document pairs into four categories:
\begin{itemize}
    \item \textbf{Q$\in$Train, D$\in$Train}: Both query and document have appeared in the train set,
    \item \textbf{Q$\notin$Train, D$\in$Train}: Query has not been seen in the train set, but the relevant document has been seen in the train set,
    \item \textbf{Q$\in$Train, D$\notin$Train}: Query has been seen in the train set, but the document is a new web page that has not been seen in the train set,
    \item \textbf{Q$\notin$Train, D$\notin$Train}: Both query and document are novel content which have never been seen in the train set.
\end{itemize}
We can see from the table \ref{tab:qc_distribution} that 82\% of query-document pairs are novel content in the test set which have not been seen in the train set. Therefore, \sysname{} dataset is capable of offering effective assessments of models based on memory capacity and generalizability by dividing the test set into four categories for a more detailed comparison.

\subsection{New Challenges Raised by \sysname{}}
Based on the \sysname{} datasets, we raise three challenge tasks in large embedding model and retrieval system design.
\subsubsection{Large-scale Embedding Model Challenge} 
As introduced before, the large-scale web data volume requires large embedding models to guarantee sufficient knowledge coverage. It requires balancing the following two goals: good model generalization ability and efficient train/inference speed.

\subsubsection{Embedding Retrieval Algorithm Challenge} Embedding models need to co-work with the embedding retrieval system to serve a web scale dataset. In this challenge, we take the embedding vectors generated by our best baseline model as the ANN vector set. The goal of this challenge is to call for ANN algorithm innovations to minimize the accuracy gap between approximate search and brute-force search while still preserving good system performance.

\subsubsection{End-to-end Retrieval System Challenge}
In the web scenario, the result quality and system performance of the end-to-end retrieval system are the most important metrics in comparing different solutions. This challenge task encourages any kind of solutions, including an embedding model plus ANN system~\cite{lu2020twinbert}, inverted index solution~\cite{dai2019context, Zhuang2021DeepQL, bevilacqua2022autoregressive}, hybrid solution~\cite{seo2019real,guo2022semantic,lassance2023naver}, neural indexer~\cite{wang2022neural, tay2022transformer}, and large language model~\cite{touvron2023llama} etc.     

\begin{table*}[!htbp]
  \centering
  \caption{Result quality of baseline models}
  \begin{tabular}{|l|c|c|c|c|c|c|c|c|c|c|}
    \hline
    \multirow{2}{*} {Baselines} & \multicolumn{6}{c|}{\sysname{}} & \multicolumn{2}{c|}{NQ} & \multicolumn{2}{c|}{MS MARCO Passage} \\
    \cline{2-11}
    & MRR@10 & recall@1 & recall@5 & recall@10 & recall@20 & recall@100 & recall@20 & recall@100 & recall@50 & recall@1K \\
    \hline
    DPR & 0.542 & 45.12\% & 66.04\% & 72.10\% & 76.80\% & 87.54\%     & 78.4\% & 85.4\% & - & - \\
    ANCE & 0.633 & 54.18\% & 75.53\% & 80.53\% & 84.17\% & 91.17\%    & 81.9\% & 87.5\% & 81.1\% & 95.9\%\\
    SimANS & \textbf{0.649} & \textbf{55.86\%} & \textbf{76.84\%} & \textbf{81.78\%} & \textbf{85.23\%} & \textbf{91.98\%}     & \textbf{86.2\%} & \textbf{90.3\%} & \textbf{88.7\%} & \textbf{98.7\%} \\
    \hline
  \end{tabular}
  \label{tab:embedding_models}
\end{table*}

\begin{table*}[]
  \centering
  \caption{Performance of ANN baselines}
  \begin{tabular}{|l|c|c|c|c|c|c|c|}
    \hline
    Baselines & ANN recall@1 & ANN recall@10 & ANN recall@100 & QPS & P50 latency & P90 latency & P99 latency  \\
    \hline
    SPANN & 87.97\% & 80.55\% & 69.84\% & 625 & 10.411 ms & 10.873 ms & 11.334 ms \\
    DiskANN & 91.46\% & 87.07\% & 69.73\% & 2691 & 21.968 ms & 37.841 & 69.462 ms \\
    \hline
  \end{tabular}
  \label{tab:ann}
\end{table*}

\begin{table*}[]
  \centering
  \caption{Result quality of baseline systems}
  \begin{tabular}{|l|c|c|c|c|c|c|}
    \hline
    Baselines & MRR@10 & recall@1 & recall@5 & recall@10 & recall@20 & recall@100 \\
    \hline
    Elasticsearch BM25 & 0.296 & 22.30\% & 39.04\% & 46.00\% & 52.42\% & 63.87\% \\
    DPR & 0.467 & 39.21\% & 56.66\% & 61.27\% & 64.69\% & 70.28\% \\
    ANCE & 0.580 & 49.87\% & 68.59\% & 72.94\% & 75.86\% & 80.18\% \\
    SimANS & 0.585 & 50.63\% & 68.79\% & 73.14\% & 75.85\% & 79.82\% \\
    \hline
  \end{tabular}
  \label{tab:e2e_embedding_models}
\end{table*}

\begin{table*}[]
  \centering
  \caption{System Performance of Baseline models}
  \begin{tabular}{|l|c|c|c|c|c|c|}
    \hline
    Baselines & QPS & P50 latency & P90 latency & P99 latency   \\
    \hline
    Elasticsearch BM25 & 149  & 312.025 ms & 1065.141 ms & 3745.546 ms \\
    DPR/ANCE/SimANS & 625  & 21.924 ms & 23.017 ms & 34.217 ms \\
    
    \hline
  \end{tabular}
  \label{tab:e2e_embedding_models_perf}
\end{table*}

\begin{figure*}
\centering
\subfigure[Documents]{
    \begin{minipage}[b]{.3\textwidth}
    \label{fig:doc_d}
    \includegraphics[width=1\linewidth]{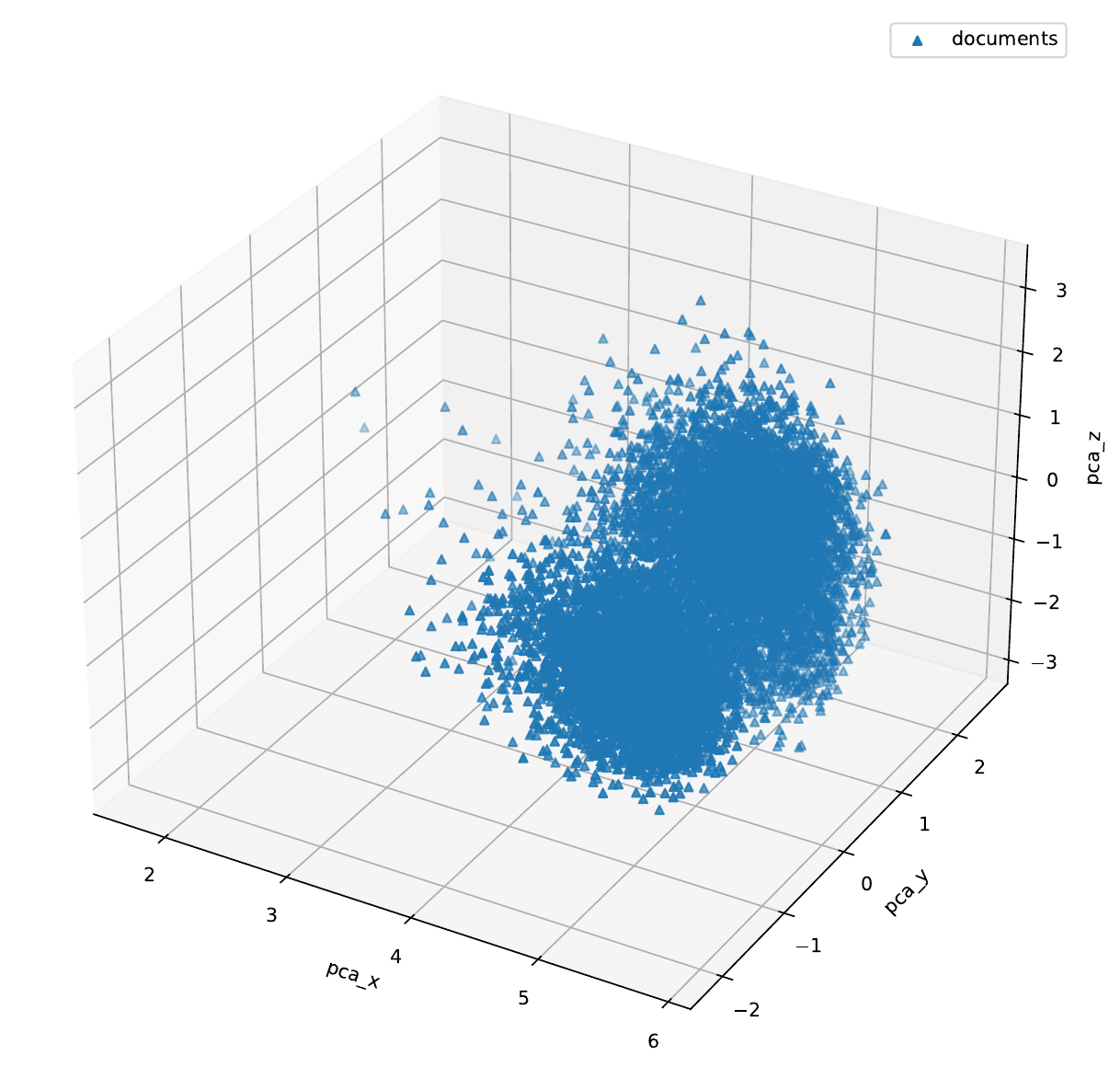}
    \end{minipage}
}
\subfigure[Queries]{
    \begin{minipage}[b]{.3\textwidth}
    \label{fig:query_d}
     \includegraphics[width=1\linewidth]{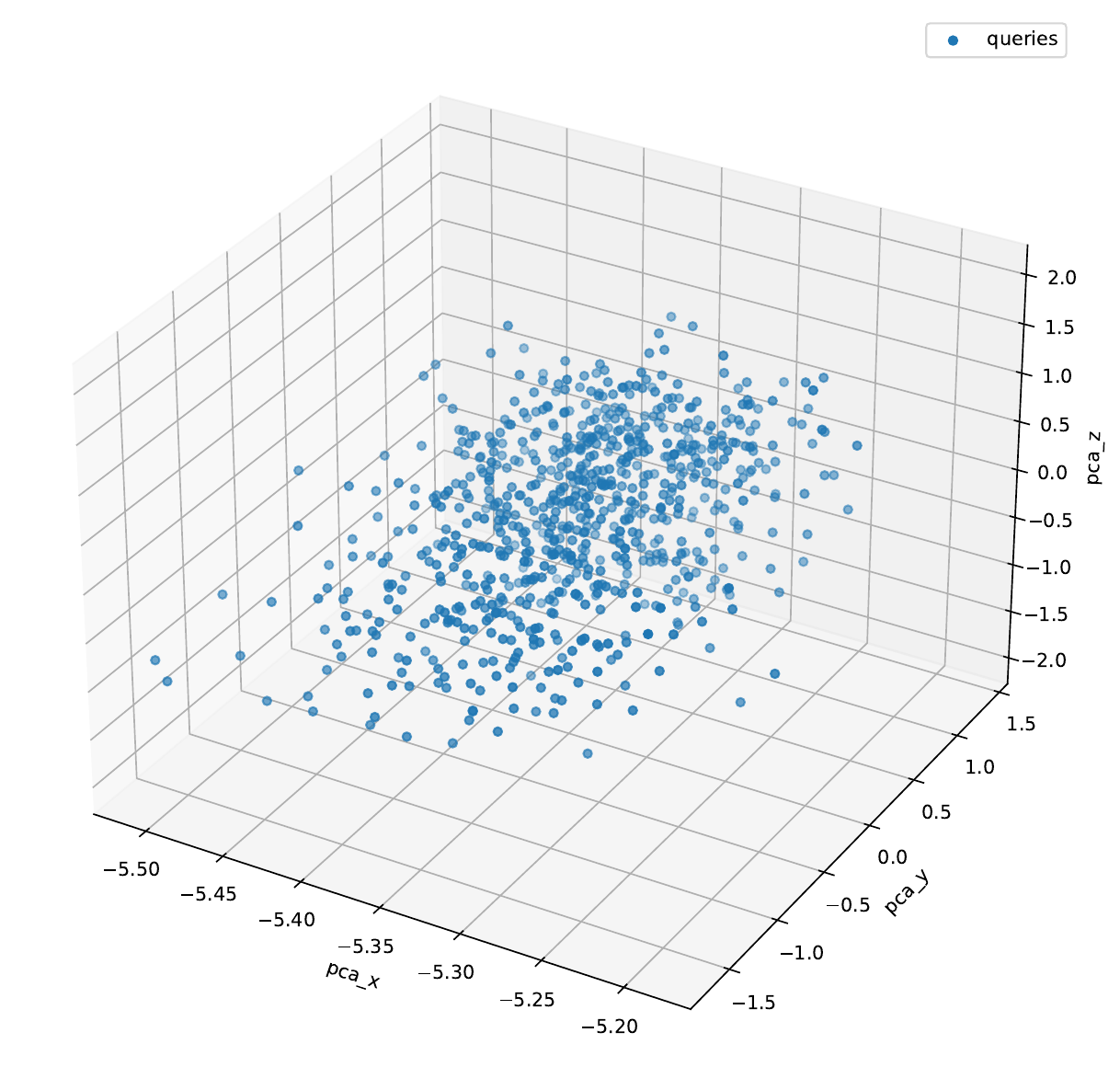}
     \end{minipage}
}
\subfigure[Documents \& Queries]{
    \begin{minipage}[b]{.3\textwidth}
      \label{fig:query_doc_d}
    \includegraphics[width=1\linewidth]{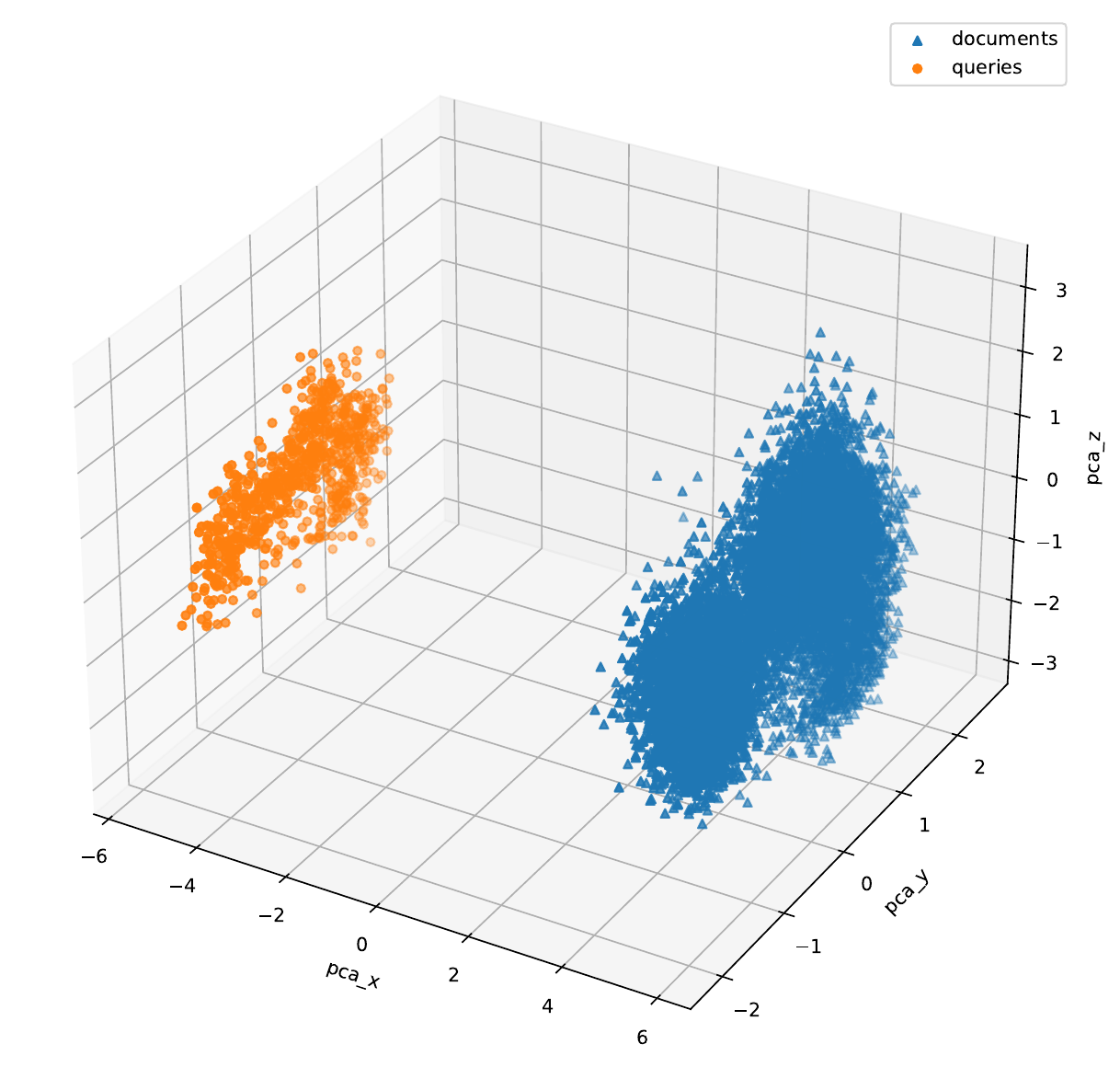}
    \end{minipage}
}
\caption{
The large gap between embedding vector distribution of documents and queries in SimANS.
}
\label{fig:dd}
\end{figure*}

\section{Benchmark Results}

In this section, we provide initial benchmark results for some state-of-the-art embedding models, ANN algorithms, and popular retrieval systems on the \sysname{} 100M dataset as baselines. For the 10B dataset, we leave it for open exploration.

\subsection{Environment Setup}
We use the Azure Standard\_ND96asr\_v4 Virtual Machine for model training and performance testing. It contains 96 vCPU cores, 900 GB memory, 8 A100 40GB GPUs with NVLink 3.0.

\subsection{Baseline Methods}

\subsubsection{State-of-the-art Embedding Models}
We take the following three state-of-the-art multi-lingual models as the initial baseline models:
\begin{itemize}
    \item DPR~\cite{karpukhin2020dense} is based on a BERT pre-trained model and a dual-encoder architecture, whose embedding is optimized for maximizing the inner product score of a query and its relevant passage. The negative training examples are selected by BM25 retrieved documents.
    \item ANCE~\cite{xiong2020approximate} improves embedding-based retrieval performance by selecting hard negative training examples from the entire document set using an asynchronously updated approximate nearest neighbor index.
    \item SimANS~\cite{zhou2022simans} avoids over indexing on false negatives by selecting ambiguous samples rather than the hardest ones.
\end{itemize}
We aspire for the \sysname{} dataset to establish itself as a new standard benchmark for retrieval, enticing more baseline models to assess and experiment with it.

\subsubsection{State-of-the-art ANN Algorithms}
For embedding retrieval algorithms, we choose the state-of-the-art disk-based ANN algorithms DiskANN~\cite{jayaram2019diskann} and SPANN~\cite{chen2021spann} as the baselines. 
DiskANN is the first disk-based ANN algorithm that can effectively serve billion-scale vector search with low resource cost. It adopts the neighborhood graph solution which stores the graph and the full-precision vectors on the disk, while putting compressed vectors (e.g., through Product Quantization~\cite{jegou2010product}) and some pivot points in memory. During the search, a query follows the best-first traversal principle to start the search from a fixed point in the graph.

SPANN adopts a hierarchical balanced clustering technique to fast partition the whole dataset into a large number of posting lists which are kept in the disk. It only stores the centroids of the posting lists in memory. To speed up the search, SPANN leverages a memory index to fast navigate the query to closest centroids, and then loads the corresponding posting lists from disk into memory for further fine-grained search. With several techniques like postings expansion with a maximum length constraint, boundary vector replication and query-aware dynamic pruning, it achieves state-of-the-art performance in multiple billion-scale datasets in terms of memory cost, result quality, and search latency.

\subsubsection{End-to-end Retrieval Systems}
BM25~\cite{robertson1994some} is the most widely used ranking function in the web information retrieval area to estimate the relevance score of a document given a query. It ranks a set of documents based on a probabilistic retrieval framework and has been integrated into the Elasticsearch system to serve all kinds of search scenarios.

\subsection{Evaluation Metrics}
We evaluate all the baselines on both result quality and system performance aspects. For result quality, we take Mean Reciprocal Rank (MRR) and recall as the evaluation metrics:
\begin{itemize}
    \item MRR: the average of the multiplicative inverse of the rank of the first correct result, which is widely used for evaluating the model quality.
    \item Recall: the average percentage of ground truth items recalled during the search. For the embedding model challenge and the end-to-end retrieval system challenge, we use our test query-document labels as the ground truth. For the embedding retrieval algorithm challenge, we use the brute-force vector search results as the ground truth (ANN recall) to evaluate the ANN algorithm performance.
\end{itemize}

For system performance, we evaluate the following metrics under limited resource cost to align with industry scenarios:
\begin{itemize}
    \item Throughput: All queries are provided at once, and we measure the wall clock time between the ingestion of the vectors and when all the results are output using all the threads in a machine. Then the throughput is calculated as the processed queries per second (QPS).
    \item Latency: we measure the 50, 90 and 99 percentile query latency at certain QPS.
\end{itemize}

\subsection{Evaluation of Embedding Models}
\label{sec:embedding_model_eval}
In this experiment, we measure the MRR and recall of all the baseline embedding models. From the result, we can see that SimANS with the ambiguous samples as negative training examples performs the best on \sysname{} 100M dataset. The ranking of the baseline models is aligned with the model evolution trend in the literature. Nonetheless, when compared with the evaluation results in Natural Question (NQ)~\cite{nq} and MS MARCO Passage Ranking~\cite{msmarco}, the gap in performance between ANCE and SimANS in \sysname{} becomes less significant.

We also evaluate the system performance for the three baseline embedding models. Since they use the same model architecture and the same number of parameters, their serving time cost is similar. At the peak of 698 QPS, the latency percentiles of 50, 90, and 99 are 9.896 ms, 10.018 ms, and 11.430 ms, respectively. 

\subsection{Evaluation of ANN Algorithms}
In this experiment, we evaluate the ANN performance with vectors generated by the best baseline model. We build both DiskANN and SPANN indices and evaluate both their serving performance and result quality. Here we only focus on evaluating the gap between ANN and KNN. Therefore, we use the brute-force search results as the ground truth to measure recall. Table~\ref{tab:ann} summarizes the recall and system performance of the two baselines. From the results, we can see that it is difficult to achieve high recall when the number of return results K is large. One of the reasons is that the distributions of queries and documents are highly-skewed and far away from each other (see figure~\ref{fig:dd}). We also observe this phenomenon in DPR and ANCE embeddings.

\subsection{Evaluation of End-to-end Performance}
\label{sec:end-to-end}
In this section, we evaluate the end-to-end performance of the three baseline embedding models plus SPANN index and the widely-used Elasticsearch BM25 solution. Table~\ref{tab:e2e_embedding_models} and table~\ref{tab:e2e_embedding_models_perf} demonstrate the result quality and system performance of all these baseline systems, respectively. Compared with table~\ref{tab:embedding_models}, we can see that after using the ANN index, the final result quality drops a lot. For example, the metric recall@100 drops more than 10 points for all baseline models. There exists large quality gaps between the ANN and KNN results (see table~\ref{tab:ann}). Moreover, we notice that using the ANN index will change the model ranking trend. SimANS achieves the best results for all the result quality metrics with brute-force search. However, when using the SPANN index, it performs worse than ANCE in recall@20 and recall@100. We further analyze the phenomenon in detail and find that SimANS has a larger gap between the average distance of query to the top100 documents relative to the average distance of document to the top100 documents than ANCE. The gap in SimANS and ANCE are 103.35 and 73.29, respectively. This will cause inaccurate distance bound estimation for a query to the neighbors of a document. As a result, ANN cannot perform well because it relies on distance estimated according to the triangle inequality. Both result quality and system performance results of the end-to-end evaluation call for more innovations on the end-to-end retrieval system design.

\section{Potential Biases and Limitations}
As discussed in section~\ref{sec:lang_analysis}, The language distribution of documents and queries in the web scenario is high-skewed. This will lead to language bias on data and models. ClueWeb22~\cite{callan2012lemur} demonstrates that there also exists topic distribution skew in the web scenario. Therefore, domain bias also may happen in data and models.

To protect user privacy and content health, we remove queries that are rarely triggered (triggered by less than K users, where K is a high value), contain personally identifiable information, offensive content, adult content and queries that have no click connection to the ClueWeb22 document set. As a result, the query distribution is slightly different from the real web query distribution.

\section{Future Work and Conclusions}
\sysname{} is the first web dataset that effectively meets the criteria of being large, real, and rich in terms of data quality. It is composed of large-scale web pages and query-document labels sourced from a commercial search engine, retaining rich information about the web pages that is widely employed in industry. The retrieval benchmark offered by \sysname{} comprises three challenging tasks that require innovation in both the areas of machine learning and information retrieval system research. We hope \sysname{} can serve as a benchmark for modern web-scale information retrieval, facilitating future research and innovation in diverse directions. 

\bibliographystyle{ACM-Reference-Format}
\bibliography{sample}


\begin{thebibliography}{58}


\ifx \showCODEN    \undefined \def \showCODEN     #1{\unskip}     \fi
\ifx \showDOI      \undefined \def \showDOI       #1{#1}\fi
\ifx \showISBNx    \undefined \def \showISBNx     #1{\unskip}     \fi
\ifx \showISBNxiii \undefined \def \showISBNxiii  #1{\unskip}     \fi
\ifx \showISSN     \undefined \def \showISSN      #1{\unskip}     \fi
\ifx \showLCCN     \undefined \def \showLCCN      #1{\unskip}     \fi
\ifx \shownote     \undefined \def \shownote      #1{#1}          \fi
\ifx \showarticletitle \undefined \def \showarticletitle #1{#1}   \fi
\ifx \showURL      \undefined \def \showURL       {\relax}        \fi
\providecommand\bibfield[2]{#2}
\providecommand\bibinfo[2]{#2}
\providecommand\natexlab[1]{#1}
\providecommand\showeprint[2][]{arXiv:#2}

\bibitem[big({[n.\,d.]})]%
        {bigann}
 \bibinfo{year}{[n.\,d.]}\natexlab{}.
\newblock \bibinfo{title}{Billion-scale ANNS Benchmarks}.
\newblock \bibinfo{howpublished}{\url{https://big-ann-benchmarks.com/}}.
\newblock


\bibitem[com({[n.\,d.]})]%
        {commoncrawl}
 \bibinfo{year}{[n.\,d.]}\natexlab{}.
\newblock \bibinfo{title}{Common Crawl}.
\newblock
\newblock


\bibitem[Rob({[n.\,d.]})]%
        {Robust04}
 \bibinfo{year}{[n.\,d.]}\natexlab{}.
\newblock \bibinfo{title}{Robust04}.
\newblock
  \bibinfo{howpublished}{\url{https://trec.nist.gov/data/robust/04.guidelines.html}}.
\newblock


\bibitem[Aum{\"u}ller et~al\mbox{.}(2017)]%
        {aumuller2017ann}
\bibfield{author}{\bibinfo{person}{Martin Aum{\"u}ller}, \bibinfo{person}{Erik
  Bernhardsson}, {and} \bibinfo{person}{Alexander Faithfull}.}
  \bibinfo{year}{2017}\natexlab{}.
\newblock \showarticletitle{ANN-benchmarks: A benchmarking tool for approximate
  nearest neighbor algorithms}. In \bibinfo{booktitle}{\emph{International
  conference on similarity search and applications}}. Springer,
  \bibinfo{pages}{34--49}.
\newblock


\bibitem[Babenko and Lempitsky(2014)]%
        {babenko2014inverted}
\bibfield{author}{\bibinfo{person}{Artem Babenko} {and} \bibinfo{person}{Victor
  Lempitsky}.} \bibinfo{year}{2014}\natexlab{}.
\newblock \showarticletitle{The inverted multi-index}.
\newblock \bibinfo{journal}{\emph{IEEE transactions on pattern analysis and
  machine intelligence}} \bibinfo{volume}{37}, \bibinfo{number}{6}
  (\bibinfo{year}{2014}), \bibinfo{pages}{1247--1260}.
\newblock


\bibitem[Babenko and Lempitsky(2016)]%
        {gnoimi2016efficient}
\bibfield{author}{\bibinfo{person}{Artem Babenko} {and} \bibinfo{person}{Victor
  Lempitsky}.} \bibinfo{year}{2016}\natexlab{}.
\newblock \showarticletitle{Efficient indexing of billion-scale datasets of
  deep descriptors}. In \bibinfo{booktitle}{\emph{Proceedings of the IEEE
  Conference on Computer Vision and Pattern Recognition (CVPR)}}.
  \bibinfo{pages}{2055--2063}.
\newblock


\bibitem[Baranchuk et~al\mbox{.}(2018)]%
        {baranchuk2018revisiting}
\bibfield{author}{\bibinfo{person}{Dmitry Baranchuk}, \bibinfo{person}{Artem
  Babenko}, {and} \bibinfo{person}{Yury Malkov}.}
  \bibinfo{year}{2018}\natexlab{}.
\newblock \showarticletitle{Revisiting the inverted indices for billion-scale
  approximate nearest neighbors}. In \bibinfo{booktitle}{\emph{Proceedings of
  the European Conference on Computer Vision (ECCV)}}.
  \bibinfo{pages}{202--216}.
\newblock


\bibitem[Bevilacqua et~al\mbox{.}(2022)]%
        {bevilacqua2022autoregressive}
\bibfield{author}{\bibinfo{person}{Michele Bevilacqua},
  \bibinfo{person}{Giuseppe Ottaviano}, \bibinfo{person}{Patrick Lewis},
  \bibinfo{person}{Scott Yih}, \bibinfo{person}{Sebastian Riedel}, {and}
  \bibinfo{person}{Fabio Petroni}.} \bibinfo{year}{2022}\natexlab{}.
\newblock \showarticletitle{Autoregressive search engines: Generating
  substrings as document identifiers}.
\newblock \bibinfo{journal}{\emph{Advances in Neural Information Processing
  Systems}}  \bibinfo{volume}{35} (\bibinfo{year}{2022}),
  \bibinfo{pages}{31668--31683}.
\newblock


\bibitem[Callan(2012)]%
        {callan2012lemur}
\bibfield{author}{\bibinfo{person}{Jamie Callan}.}
  \bibinfo{year}{2012}\natexlab{}.
\newblock \showarticletitle{The lemur project and its clueweb12 dataset}. In
  \bibinfo{booktitle}{\emph{Invited talk at the SIGIR 2012 Workshop on
  Open-Source Information Retrieval}}.
\newblock


\bibitem[Chen et~al\mbox{.}(2024)]%
        {chen2024bge}
\bibfield{author}{\bibinfo{person}{Jianlv Chen}, \bibinfo{person}{Shitao Xiao},
  \bibinfo{person}{Peitian Zhang}, \bibinfo{person}{Kun Luo},
  \bibinfo{person}{Defu Lian}, {and} \bibinfo{person}{Zheng Liu}.}
  \bibinfo{year}{2024}\natexlab{}.
\newblock \showarticletitle{Bge m3-embedding: Multi-lingual,
  multi-functionality, multi-granularity text embeddings through self-knowledge
  distillation}.
\newblock \bibinfo{journal}{\emph{arXiv preprint arXiv:2402.03216}}
  (\bibinfo{year}{2024}).
\newblock


\bibitem[Chen et~al\mbox{.}(2021)]%
        {chen2021spann}
\bibfield{author}{\bibinfo{person}{Qi Chen}, \bibinfo{person}{Bing Zhao},
  \bibinfo{person}{Haidong Wang}, \bibinfo{person}{Mingqin Li},
  \bibinfo{person}{Chuanjie Liu}, \bibinfo{person}{Zengzhong Li},
  \bibinfo{person}{Mao Yang}, {and} \bibinfo{person}{Jingdong Wang}.}
  \bibinfo{year}{2021}\natexlab{}.
\newblock \showarticletitle{SPANN: Highly-efficient Billion-scale Approximate
  Nearest Neighborhood Search}.
\newblock \bibinfo{journal}{\emph{Advances in Neural Information Processing
  Systems}}  \bibinfo{volume}{34} (\bibinfo{year}{2021}),
  \bibinfo{pages}{5199--5212}.
\newblock


\bibitem[Clarke et~al\mbox{.}(2004)]%
        {Clarke2004TrecTerabyte}
\bibfield{author}{\bibinfo{person}{Charles Clarke}, \bibinfo{person}{Nick
  Craswell}, {and} \bibinfo{person}{Ian Soboroff}.}
  \bibinfo{year}{2004}\natexlab{}.
\newblock \showarticletitle{Overview of the TREC 2004 Terabyte Track}. In
  \bibinfo{booktitle}{\emph{TREC}}.
\newblock


\bibitem[Clarke et~al\mbox{.}(2009)]%
        {clarke2009overview}
\bibfield{author}{\bibinfo{person}{Charles~LA Clarke}, \bibinfo{person}{Nick
  Craswell}, {and} \bibinfo{person}{Ian Soboroff}.}
  \bibinfo{year}{2009}\natexlab{}.
\newblock \showarticletitle{Overview of the TREC 2009 Web Track.}. In
  \bibinfo{booktitle}{\emph{Trec}}, Vol.~\bibinfo{volume}{9}.
  \bibinfo{pages}{20--29}.
\newblock


\bibitem[Craswell et~al\mbox{.}(2020)]%
        {craswell2020orcas}
\bibfield{author}{\bibinfo{person}{Nick Craswell}, \bibinfo{person}{Daniel
  Campos}, \bibinfo{person}{Bhaskar Mitra}, \bibinfo{person}{Emine Yilmaz},
  {and} \bibinfo{person}{Bodo Billerbeck}.} \bibinfo{year}{2020}\natexlab{}.
\newblock \showarticletitle{ORCAS: 20 million clicked query-document pairs for
  analyzing search}. In \bibinfo{booktitle}{\emph{Proceedings of the 29th ACM
  International Conference on Information \& Knowledge Management}}.
  \bibinfo{pages}{2983--2989}.
\newblock


\bibitem[Dai and Callan(2019)]%
        {dai2019context}
\bibfield{author}{\bibinfo{person}{Zhuyun Dai} {and} \bibinfo{person}{Jamie
  Callan}.} \bibinfo{year}{2019}\natexlab{}.
\newblock \showarticletitle{Context-aware sentence/passage term importance
  estimation for first stage retrieval}.
\newblock \bibinfo{journal}{\emph{arXiv preprint arXiv:1910.10687}}
  (\bibinfo{year}{2019}).
\newblock


\bibitem[Devlin et~al\mbox{.}(2018)]%
        {devlin2018bert}
\bibfield{author}{\bibinfo{person}{Jacob Devlin}, \bibinfo{person}{Ming-Wei
  Chang}, \bibinfo{person}{Kenton Lee}, {and} \bibinfo{person}{Kristina
  Toutanova}.} \bibinfo{year}{2018}\natexlab{}.
\newblock \showarticletitle{Bert: Pre-training of deep bidirectional
  transformers for language understanding}.
\newblock \bibinfo{journal}{\emph{arXiv preprint arXiv:1810.04805}}
  (\bibinfo{year}{2018}).
\newblock


\bibitem[Gao and Callan(2022)]%
        {gao2022unsupervised}
\bibfield{author}{\bibinfo{person}{Luyu Gao} {and} \bibinfo{person}{Jamie
  Callan}.} \bibinfo{year}{2022}\natexlab{}.
\newblock \showarticletitle{Unsupervised Corpus Aware Language Model
  Pre-training for Dense Passage Retrieval}. In
  \bibinfo{booktitle}{\emph{Proceedings of the 60th Annual Meeting of the
  Association for Computational Linguistics (Volume 1: Long Papers)}}.
  \bibinfo{pages}{2843--2853}.
\newblock


\bibitem[Guo et~al\mbox{.}(2022)]%
        {guo2022semantic}
\bibfield{author}{\bibinfo{person}{Jiafeng Guo}, \bibinfo{person}{Yinqiong
  Cai}, \bibinfo{person}{Yixing Fan}, \bibinfo{person}{Fei Sun},
  \bibinfo{person}{Ruqing Zhang}, {and} \bibinfo{person}{Xueqi Cheng}.}
  \bibinfo{year}{2022}\natexlab{}.
\newblock \showarticletitle{Semantic models for the first-stage retrieval: A
  comprehensive review}.
\newblock \bibinfo{journal}{\emph{ACM Transactions on Information Systems
  (TOIS)}} \bibinfo{volume}{40}, \bibinfo{number}{4} (\bibinfo{year}{2022}),
  \bibinfo{pages}{1--42}.
\newblock


\bibitem[Guo et~al\mbox{.}(2020)]%
        {scann2020}
\bibfield{author}{\bibinfo{person}{Ruiqi Guo}, \bibinfo{person}{Philip Sun},
  \bibinfo{person}{Erik Lindgren}, \bibinfo{person}{Quan Geng},
  \bibinfo{person}{David Simcha}, \bibinfo{person}{Felix Chern}, {and}
  \bibinfo{person}{Sanjiv Kumar}.} \bibinfo{year}{2020}\natexlab{}.
\newblock \showarticletitle{Accelerating Large-Scale Inference with Anisotropic
  Vector Quantization}. In \bibinfo{booktitle}{\emph{Proceedings of the 37th
  International Conference on Machine Learning (ICML)}}.
  \bibinfo{pages}{3887--3896}.
\newblock


\bibitem[Hu et~al\mbox{.}(2014)]%
        {hu2014convolutional}
\bibfield{author}{\bibinfo{person}{Baotian Hu}, \bibinfo{person}{Zhengdong Lu},
  \bibinfo{person}{Hang Li}, {and} \bibinfo{person}{Qingcai Chen}.}
  \bibinfo{year}{2014}\natexlab{}.
\newblock \showarticletitle{Convolutional neural network architectures for
  matching natural language sentences}.
\newblock \bibinfo{journal}{\emph{Advances in neural information processing
  systems}}  \bibinfo{volume}{27} (\bibinfo{year}{2014}).
\newblock


\bibitem[Huang et~al\mbox{.}(2013)]%
        {huang2013learning}
\bibfield{author}{\bibinfo{person}{Po-Sen Huang}, \bibinfo{person}{Xiaodong
  He}, \bibinfo{person}{Jianfeng Gao}, \bibinfo{person}{Li Deng},
  \bibinfo{person}{Alex Acero}, {and} \bibinfo{person}{Larry Heck}.}
  \bibinfo{year}{2013}\natexlab{}.
\newblock \showarticletitle{Learning deep structured semantic models for web
  search using clickthrough data}. In \bibinfo{booktitle}{\emph{Proceedings of
  the 22nd ACM international conference on Information \& Knowledge
  Management}}. \bibinfo{pages}{2333--2338}.
\newblock


\bibitem[Jayaram~Subramanya et~al\mbox{.}(2019)]%
        {jayaram2019diskann}
\bibfield{author}{\bibinfo{person}{Suhas Jayaram~Subramanya},
  \bibinfo{person}{Fnu Devvrit}, \bibinfo{person}{Harsha~Vardhan Simhadri},
  \bibinfo{person}{Ravishankar Krishnawamy}, {and} \bibinfo{person}{Rohan
  Kadekodi}.} \bibinfo{year}{2019}\natexlab{}.
\newblock \showarticletitle{Diskann: Fast accurate billion-point nearest
  neighbor search on a single node}.
\newblock \bibinfo{journal}{\emph{Advances in Neural Information Processing
  Systems}}  \bibinfo{volume}{32} (\bibinfo{year}{2019}).
\newblock


\bibitem[Jegou et~al\mbox{.}(2010)]%
        {jegou2010product}
\bibfield{author}{\bibinfo{person}{Herve Jegou}, \bibinfo{person}{Matthijs
  Douze}, {and} \bibinfo{person}{Cordelia Schmid}.}
  \bibinfo{year}{2010}\natexlab{}.
\newblock \showarticletitle{Product quantization for nearest neighbor search}.
\newblock \bibinfo{journal}{\emph{IEEE transactions on pattern analysis and
  machine intelligence}} \bibinfo{volume}{33}, \bibinfo{number}{1}
  (\bibinfo{year}{2010}), \bibinfo{pages}{117--128}.
\newblock


\bibitem[J{\'e}gou et~al\mbox{.}(2011)]%
        {jegou2011searching}
\bibfield{author}{\bibinfo{person}{Herv{\'e} J{\'e}gou},
  \bibinfo{person}{Romain Tavenard}, \bibinfo{person}{Matthijs Douze}, {and}
  \bibinfo{person}{Laurent Amsaleg}.} \bibinfo{year}{2011}\natexlab{}.
\newblock \showarticletitle{Searching in one billion vectors: re-rank with
  source coding}. In \bibinfo{booktitle}{\emph{Proceedings of the IEEE
  International Conference on Acoustics, Speech and Signal Processing
  (ICASSP)}}. \bibinfo{pages}{861--864}.
\newblock


\bibitem[Johnson et~al\mbox{.}(2019)]%
        {faiss17}
\bibfield{author}{\bibinfo{person}{Jeff Johnson}, \bibinfo{person}{Matthijs
  Douze}, {and} \bibinfo{person}{Herv{\'e} J{\'e}gou}.}
  \bibinfo{year}{2019}\natexlab{}.
\newblock \showarticletitle{Billion-scale similarity search with GPUs}.
\newblock \bibinfo{journal}{\emph{IEEE Transactions on Big Data}}
  (\bibinfo{year}{2019}).
\newblock


\bibitem[Kalantidis and Avrithis(2014)]%
        {lopq2014locally}
\bibfield{author}{\bibinfo{person}{Yannis Kalantidis} {and}
  \bibinfo{person}{Yannis Avrithis}.} \bibinfo{year}{2014}\natexlab{}.
\newblock \showarticletitle{Locally optimized product quantization for
  approximate nearest neighbor search}. In
  \bibinfo{booktitle}{\emph{Proceedings of the IEEE Conference on Computer
  Vision and Pattern Recognition (CVPR)}}. \bibinfo{pages}{2321--2328}.
\newblock


\bibitem[Karpukhin et~al\mbox{.}(2020)]%
        {karpukhin2020dense}
\bibfield{author}{\bibinfo{person}{Vladimir Karpukhin}, \bibinfo{person}{Barlas
  O{\u{g}}uz}, \bibinfo{person}{Sewon Min}, \bibinfo{person}{Patrick Lewis},
  \bibinfo{person}{Ledell Wu}, \bibinfo{person}{Sergey Edunov},
  \bibinfo{person}{Danqi Chen}, {and} \bibinfo{person}{Wen-tau Yih}.}
  \bibinfo{year}{2020}\natexlab{}.
\newblock \showarticletitle{Dense passage retrieval for open-domain question
  answering}.
\newblock \bibinfo{journal}{\emph{arXiv preprint arXiv:2004.04906}}
  (\bibinfo{year}{2020}).
\newblock


\bibitem[Kwiatkowski et~al\mbox{.}(2019)]%
        {nq}
\bibfield{author}{\bibinfo{person}{Tom Kwiatkowski},
  \bibinfo{person}{Jennimaria Palomaki}, \bibinfo{person}{Olivia Redfield},
  \bibinfo{person}{Michael Collins}, \bibinfo{person}{Ankur Parikh},
  \bibinfo{person}{Chris Alberti}, \bibinfo{person}{Danielle Epstein},
  \bibinfo{person}{Illia Polosukhin}, \bibinfo{person}{Jacob Devlin},
  \bibinfo{person}{Kenton Lee}, {et~al\mbox{.}}}
  \bibinfo{year}{2019}\natexlab{}.
\newblock \showarticletitle{Natural questions: a benchmark for question
  answering research}.
\newblock \bibinfo{journal}{\emph{Transactions of the Association for
  Computational Linguistics}}  \bibinfo{volume}{7} (\bibinfo{year}{2019}),
  \bibinfo{pages}{453--466}.
\newblock


\bibitem[Lassance and Clinchant(2023)]%
        {lassance2023naver}
\bibfield{author}{\bibinfo{person}{Carlos Lassance} {and}
  \bibinfo{person}{St{\'e}phane Clinchant}.} \bibinfo{year}{2023}\natexlab{}.
\newblock \showarticletitle{Naver Labs Europe (SPLADE)@ TREC Deep Learning
  2022}.
\newblock \bibinfo{journal}{\emph{arXiv preprint arXiv:2302.12574}}
  (\bibinfo{year}{2023}).
\newblock


\bibitem[Lewis et~al\mbox{.}(2021)]%
        {lewis2021question}
\bibfield{author}{\bibinfo{person}{Patrick Lewis}, \bibinfo{person}{Pontus
  Stenetorp}, {and} \bibinfo{person}{Sebastian Riedel}.}
  \bibinfo{year}{2021}\natexlab{}.
\newblock \showarticletitle{Question and Answer Test-Train Overlap in
  Open-Domain Question Answering Datasets}. In
  \bibinfo{booktitle}{\emph{Proceedings of the 16th Conference of the European
  Chapter of the Association for Computational Linguistics: Main Volume}}.
  \bibinfo{pages}{1000--1008}.
\newblock


\bibitem[Lu et~al\mbox{.}(2020)]%
        {lu2020twinbert}
\bibfield{author}{\bibinfo{person}{Wenhao Lu}, \bibinfo{person}{Jian Jiao},
  {and} \bibinfo{person}{Ruofei Zhang}.} \bibinfo{year}{2020}\natexlab{}.
\newblock \showarticletitle{Twinbert: Distilling knowledge to twin-structured
  compressed BERT models for large-scale retrieval}. In
  \bibinfo{booktitle}{\emph{Proceedings of the 29th ACM International
  Conference on Information \& Knowledge Management}}.
  \bibinfo{pages}{2645--2652}.
\newblock


\bibitem[microsoft(0a)]%
        {bing}
\bibfield{author}{\bibinfo{person}{microsoft}.} \bibinfo{year}{0}\natexlab{a}.
\newblock \bibinfo{title}{Bing search}.
\newblock \bibinfo{howpublished}{\url{https://www.bing.com/}}.
\newblock


\bibitem[microsoft(0b)]%
        {newbing}
\bibfield{author}{\bibinfo{person}{microsoft}.} \bibinfo{year}{0}\natexlab{b}.
\newblock \bibinfo{title}{New Bing}.
\newblock \bibinfo{howpublished}{\url{https://www.bing.com/new}}.
\newblock


\bibitem[Nakano et~al\mbox{.}(2021)]%
        {nakano2021webgpt}
\bibfield{author}{\bibinfo{person}{Reiichiro Nakano}, \bibinfo{person}{Jacob
  Hilton}, \bibinfo{person}{Suchir Balaji}, \bibinfo{person}{Jeff Wu},
  \bibinfo{person}{Long Ouyang}, \bibinfo{person}{Christina Kim},
  \bibinfo{person}{Christopher Hesse}, \bibinfo{person}{Shantanu Jain},
  \bibinfo{person}{Vineet Kosaraju}, \bibinfo{person}{William Saunders},
  {et~al\mbox{.}}} \bibinfo{year}{2021}\natexlab{}.
\newblock \showarticletitle{Webgpt: Browser-assisted question-answering with
  human feedback}.
\newblock \bibinfo{journal}{\emph{arXiv preprint arXiv:2112.09332}}
  (\bibinfo{year}{2021}).
\newblock


\bibitem[Nguyen et~al\mbox{.}(2016)]%
        {msmarco}
\bibfield{author}{\bibinfo{person}{Tri Nguyen}, \bibinfo{person}{Mir
  Rosenberg}, \bibinfo{person}{Xia Song}, \bibinfo{person}{Jianfeng Gao},
  \bibinfo{person}{Saurabh Tiwary}, \bibinfo{person}{Rangan Majumder}, {and}
  \bibinfo{person}{Li Deng}.} \bibinfo{year}{2016}\natexlab{}.
\newblock \showarticletitle{MS MARCO: A human generated machine reading
  comprehension dataset}. In \bibinfo{booktitle}{\emph{CoCo@ NIPS}}.
\newblock


\bibitem[Ouyang et~al\mbox{.}(2022)]%
        {ouyang2022training}
\bibfield{author}{\bibinfo{person}{Long Ouyang}, \bibinfo{person}{Jeffrey Wu},
  \bibinfo{person}{Xu Jiang}, \bibinfo{person}{Diogo Almeida},
  \bibinfo{person}{Carroll Wainwright}, \bibinfo{person}{Pamela Mishkin},
  \bibinfo{person}{Chong Zhang}, \bibinfo{person}{Sandhini Agarwal},
  \bibinfo{person}{Katarina Slama}, \bibinfo{person}{Alex Ray},
  {et~al\mbox{.}}} \bibinfo{year}{2022}\natexlab{}.
\newblock \showarticletitle{Training language models to follow instructions
  with human feedback}.
\newblock \bibinfo{journal}{\emph{Advances in Neural Information Processing
  Systems}}  \bibinfo{volume}{35} (\bibinfo{year}{2022}),
  \bibinfo{pages}{27730--27744}.
\newblock


\bibitem[Overwijk et~al\mbox{.}(2022)]%
        {overwijk2022clueweb22}
\bibfield{author}{\bibinfo{person}{Arnold Overwijk}, \bibinfo{person}{Chenyan
  Xiong}, {and} \bibinfo{person}{Jamie Callan}.}
  \bibinfo{year}{2022}\natexlab{}.
\newblock \showarticletitle{ClueWeb22: 10 billion web documents with rich
  information}. In \bibinfo{booktitle}{\emph{Proceedings of the 45th
  International ACM SIGIR Conference on Research and Development in Information
  Retrieval}}. \bibinfo{pages}{3360--3362}.
\newblock


\bibitem[Palangi et~al\mbox{.}(2016)]%
        {palangi2016deep}
\bibfield{author}{\bibinfo{person}{Hamid Palangi}, \bibinfo{person}{Li Deng},
  \bibinfo{person}{Yelong Shen}, \bibinfo{person}{Jianfeng Gao},
  \bibinfo{person}{Xiaodong He}, \bibinfo{person}{Jianshu Chen},
  \bibinfo{person}{Xinying Song}, {and} \bibinfo{person}{Rabab Ward}.}
  \bibinfo{year}{2016}\natexlab{}.
\newblock \showarticletitle{Deep sentence embedding using long short-term
  memory networks: Analysis and application to information retrieval}.
\newblock \bibinfo{journal}{\emph{IEEE/ACM Transactions on Audio, Speech, and
  Language Processing}} \bibinfo{volume}{24}, \bibinfo{number}{4}
  (\bibinfo{year}{2016}), \bibinfo{pages}{694--707}.
\newblock


\bibitem[Qiao et~al\mbox{.}(2019)]%
        {qiao2019understanding}
\bibfield{author}{\bibinfo{person}{Yifan Qiao}, \bibinfo{person}{Chenyan
  Xiong}, \bibinfo{person}{Zhenghao Liu}, {and} \bibinfo{person}{Zhiyuan Liu}.}
  \bibinfo{year}{2019}\natexlab{}.
\newblock \showarticletitle{Understanding the Behaviors of BERT in Ranking}.
\newblock \bibinfo{journal}{\emph{arXiv preprint arXiv:1904.07531}}
  (\bibinfo{year}{2019}).
\newblock


\bibitem[Reimers and Gurevych(2019)]%
        {reimers2019sentence}
\bibfield{author}{\bibinfo{person}{Nils Reimers} {and} \bibinfo{person}{Iryna
  Gurevych}.} \bibinfo{year}{2019}\natexlab{}.
\newblock \showarticletitle{Sentence-bert: Sentence embeddings using siamese
  bert-networks}.
\newblock \bibinfo{journal}{\emph{arXiv preprint arXiv:1908.10084}}
  (\bibinfo{year}{2019}).
\newblock


\bibitem[Ren et~al\mbox{.}(2020)]%
        {hmann2020}
\bibfield{author}{\bibinfo{person}{Jie Ren}, \bibinfo{person}{Minjia Zhang},
  {and} \bibinfo{person}{Dong Li}.} \bibinfo{year}{2020}\natexlab{}.
\newblock \showarticletitle{HM-ANN: Efficient Billion-Point Nearest Neighbor
  Search on Heterogeneous Memory}. In \bibinfo{booktitle}{\emph{In Proceedings
  of the 34th International Conference on Neural Information Processing
  Systems}}, Vol.~\bibinfo{volume}{33}.
\newblock


\bibitem[Robertson and Walker(1994)]%
        {robertson1994some}
\bibfield{author}{\bibinfo{person}{Stephen~E Robertson} {and}
  \bibinfo{person}{Steve Walker}.} \bibinfo{year}{1994}\natexlab{}.
\newblock \showarticletitle{Some simple effective approximations to the
  2-poisson model for probabilistic weighted retrieval}. In
  \bibinfo{booktitle}{\emph{SIGIR’94: Proceedings of the Seventeenth Annual
  International ACM-SIGIR Conference on Research and Development in Information
  Retrieval, organised by Dublin City University}}. Springer,
  \bibinfo{pages}{232--241}.
\newblock


\bibitem[Sasaki et~al\mbox{.}(2018)]%
        {sasaki2018cross}
\bibfield{author}{\bibinfo{person}{Shota Sasaki}, \bibinfo{person}{Shuo Sun},
  \bibinfo{person}{Shigehiko Schamoni}, \bibinfo{person}{Kevin Duh}, {and}
  \bibinfo{person}{Kentaro Inui}.} \bibinfo{year}{2018}\natexlab{}.
\newblock \showarticletitle{Cross-lingual learning-to-rank with shared
  representations}. In \bibinfo{booktitle}{\emph{Proceedings of the 2018
  Conference of the North American Chapter of the Association for Computational
  Linguistics: Human Language Technologies, Volume 2 (Short Papers)}}.
  \bibinfo{pages}{458--463}.
\newblock


\bibitem[Seo et~al\mbox{.}(2019)]%
        {seo2019real}
\bibfield{author}{\bibinfo{person}{Minjoon Seo}, \bibinfo{person}{Jinhyuk Lee},
  \bibinfo{person}{Tom Kwiatkowski}, \bibinfo{person}{Ankur Parikh},
  \bibinfo{person}{Ali Farhadi}, {and} \bibinfo{person}{Hannaneh Hajishirzi}.}
  \bibinfo{year}{2019}\natexlab{}.
\newblock \showarticletitle{Real-Time Open-Domain Question Answering with
  Dense-Sparse Phrase Index}. In \bibinfo{booktitle}{\emph{Proceedings of the
  57th Annual Meeting of the Association for Computational Linguistics}}.
  \bibinfo{pages}{4430--4441}.
\newblock


\bibitem[Shan et~al\mbox{.}(2021)]%
        {shan2021glow}
\bibfield{author}{\bibinfo{person}{Xuan Shan}, \bibinfo{person}{Chuanjie Liu},
  \bibinfo{person}{Yiqian Xia}, \bibinfo{person}{Qi Chen},
  \bibinfo{person}{Yusi Zhang}, \bibinfo{person}{Kaize Ding},
  \bibinfo{person}{Yaobo Liang}, \bibinfo{person}{Angen Luo}, {and}
  \bibinfo{person}{Yuxiang Luo}.} \bibinfo{year}{2021}\natexlab{}.
\newblock \showarticletitle{GLOW: Global Weighted Self-Attention Network for
  Web Search}. In \bibinfo{booktitle}{\emph{2021 IEEE International Conference
  on Big Data (Big Data)}}. IEEE, \bibinfo{pages}{519--528}.
\newblock


\bibitem[Shen et~al\mbox{.}(2014)]%
        {shen2014learning}
\bibfield{author}{\bibinfo{person}{Yelong Shen}, \bibinfo{person}{Xiaodong He},
  \bibinfo{person}{Jianfeng Gao}, \bibinfo{person}{Li Deng}, {and}
  \bibinfo{person}{Gr{\'e}goire Mesnil}.} \bibinfo{year}{2014}\natexlab{}.
\newblock \showarticletitle{Learning semantic representations using
  convolutional neural networks for web search}. In
  \bibinfo{booktitle}{\emph{Proceedings of the 23rd international conference on
  world wide web}}. \bibinfo{pages}{373--374}.
\newblock


\bibitem[Soboroff(2021)]%
        {soboroff2021overview}
\bibfield{author}{\bibinfo{person}{Ian Soboroff}.}
  \bibinfo{year}{2021}\natexlab{}.
\newblock \showarticletitle{Overview of TREC 2021}. In
  \bibinfo{booktitle}{\emph{30th Text REtrieval Conference. Gaithersburg,
  Maryland}}.
\newblock


\bibitem[Subramanya et~al\mbox{.}(2019)]%
        {jayaram2019rand}
\bibfield{author}{\bibinfo{person}{Suhas~Jayaram Subramanya},
  \bibinfo{person}{Rohan Kadekodi}, \bibinfo{person}{Ravishankar Krishaswamy},
  {and} \bibinfo{person}{Harsha~Vardhan Simhadri}.}
  \bibinfo{year}{2019}\natexlab{}.
\newblock \showarticletitle{Diskann: Fast accurate billion-point nearest
  neighbor search on a single node}. In \bibinfo{booktitle}{\emph{Proceedings
  of the 33rd International Conference on Neural Information Processing
  Systems}}. \bibinfo{pages}{13766--13776}.
\newblock


\bibitem[Tay et~al\mbox{.}(2022)]%
        {tay2022transformer}
\bibfield{author}{\bibinfo{person}{Yi Tay}, \bibinfo{person}{Vinh Tran},
  \bibinfo{person}{Mostafa Dehghani}, \bibinfo{person}{Jianmo Ni},
  \bibinfo{person}{Dara Bahri}, \bibinfo{person}{Harsh Mehta},
  \bibinfo{person}{Zhen Qin}, \bibinfo{person}{Kai Hui}, \bibinfo{person}{Zhe
  Zhao}, \bibinfo{person}{Jai Gupta}, {et~al\mbox{.}}}
  \bibinfo{year}{2022}\natexlab{}.
\newblock \showarticletitle{Transformer memory as a differentiable search
  index}.
\newblock \bibinfo{journal}{\emph{Advances in Neural Information Processing
  Systems}}  \bibinfo{volume}{35} (\bibinfo{year}{2022}),
  \bibinfo{pages}{21831--21843}.
\newblock


\bibitem[Touvron et~al\mbox{.}(2023)]%
        {touvron2023llama}
\bibfield{author}{\bibinfo{person}{Hugo Touvron}, \bibinfo{person}{Thibaut
  Lavril}, \bibinfo{person}{Gautier Izacard}, \bibinfo{person}{Xavier
  Martinet}, \bibinfo{person}{Marie-Anne Lachaux},
  \bibinfo{person}{Timoth{\'e}e Lacroix}, \bibinfo{person}{Baptiste
  Rozi{\`e}re}, \bibinfo{person}{Naman Goyal}, \bibinfo{person}{Eric Hambro},
  \bibinfo{person}{Faisal Azhar}, {et~al\mbox{.}}}
  \bibinfo{year}{2023}\natexlab{}.
\newblock \showarticletitle{Llama: Open and efficient foundation language
  models}.
\newblock \bibinfo{journal}{\emph{arXiv preprint arXiv:2302.13971}}
  (\bibinfo{year}{2023}).
\newblock


\bibitem[Wang et~al\mbox{.}(2022)]%
        {wang2022neural}
\bibfield{author}{\bibinfo{person}{Yujing Wang}, \bibinfo{person}{Yingyan Hou},
  \bibinfo{person}{Haonan Wang}, \bibinfo{person}{Ziming Miao},
  \bibinfo{person}{Shibin Wu}, \bibinfo{person}{Qi Chen},
  \bibinfo{person}{Yuqing Xia}, \bibinfo{person}{Chengmin Chi},
  \bibinfo{person}{Guoshuai Zhao}, \bibinfo{person}{Zheng Liu},
  {et~al\mbox{.}}} \bibinfo{year}{2022}\natexlab{}.
\newblock \showarticletitle{A neural corpus indexer for document retrieval}.
\newblock \bibinfo{journal}{\emph{Advances in Neural Information Processing
  Systems}}  \bibinfo{volume}{35} (\bibinfo{year}{2022}),
  \bibinfo{pages}{25600--25614}.
\newblock


\bibitem[Xiao et~al\mbox{.}(2022)]%
        {xiao2022distill}
\bibfield{author}{\bibinfo{person}{Shitao Xiao}, \bibinfo{person}{Zheng Liu},
  \bibinfo{person}{Weihao Han}, \bibinfo{person}{Jianjin Zhang},
  \bibinfo{person}{Defu Lian}, \bibinfo{person}{Yeyun Gong},
  \bibinfo{person}{Qi Chen}, \bibinfo{person}{Fan Yang}, \bibinfo{person}{Hao
  Sun}, \bibinfo{person}{Yingxia Shao}, {et~al\mbox{.}}}
  \bibinfo{year}{2022}\natexlab{}.
\newblock \showarticletitle{Distill-vq: Learning retrieval oriented vector
  quantization by distilling knowledge from dense embeddings}. In
  \bibinfo{booktitle}{\emph{Proceedings of the 45th International ACM SIGIR
  Conference on Research and Development in Information Retrieval}}.
  \bibinfo{pages}{1513--1523}.
\newblock


\bibitem[Xiao et~al\mbox{.}(2023)]%
        {xiao2023c}
\bibfield{author}{\bibinfo{person}{Shitao Xiao}, \bibinfo{person}{Zheng Liu},
  \bibinfo{person}{Peitian Zhang}, {and} \bibinfo{person}{Niklas Muennighof}.}
  \bibinfo{year}{2023}\natexlab{}.
\newblock \showarticletitle{C-pack: Packaged resources to advance general
  chinese embedding}.
\newblock \bibinfo{journal}{\emph{arXiv preprint arXiv:2309.07597}}
  (\bibinfo{year}{2023}).
\newblock


\bibitem[Xiong et~al\mbox{.}(2020)]%
        {xiong2020approximate}
\bibfield{author}{\bibinfo{person}{Lee Xiong}, \bibinfo{person}{Chenyan Xiong},
  \bibinfo{person}{Ye Li}, \bibinfo{person}{Kwok-Fung Tang},
  \bibinfo{person}{Jialin Liu}, \bibinfo{person}{Paul Bennett},
  \bibinfo{person}{Junaid Ahmed}, {and} \bibinfo{person}{Arnold Overwijk}.}
  \bibinfo{year}{2020}\natexlab{}.
\newblock \showarticletitle{Approximate nearest neighbor negative contrastive
  learning for dense text retrieval}.
\newblock \bibinfo{journal}{\emph{arXiv preprint arXiv:2007.00808}}
  (\bibinfo{year}{2020}).
\newblock


\bibitem[Xu et~al\mbox{.}(2021)]%
        {xu2021leveraging}
\bibfield{author}{\bibinfo{person}{Linlong Xu}, \bibinfo{person}{Baosong Yang},
  \bibinfo{person}{Xiaoyu Lv}, \bibinfo{person}{Tianchi Bi},
  \bibinfo{person}{Dayiheng Liu}, {and} \bibinfo{person}{Haibo Zhang}.}
  \bibinfo{year}{2021}\natexlab{}.
\newblock \showarticletitle{Leveraging Advantages of Interactive and
  Non-Interactive Models for Vector-Based Cross-Lingual Information Retrieval}.
\newblock \bibinfo{journal}{\emph{arXiv preprint arXiv:2111.01992}}
  (\bibinfo{year}{2021}).
\newblock


\bibitem[Zhan et~al\mbox{.}(2022)]%
        {zhan2022evaluating}
\bibfield{author}{\bibinfo{person}{Jingtao Zhan}, \bibinfo{person}{Xiaohui
  Xie}, \bibinfo{person}{Jiaxin Mao}, \bibinfo{person}{Yiqun Liu},
  \bibinfo{person}{Jiafeng Guo}, \bibinfo{person}{Min Zhang}, {and}
  \bibinfo{person}{Shaoping Ma}.} \bibinfo{year}{2022}\natexlab{}.
\newblock \showarticletitle{Evaluating Interpolation and Extrapolation
  Performance of Neural Retrieval Models}. In
  \bibinfo{booktitle}{\emph{Proceedings of the 31st ACM International
  Conference on Information \& Knowledge Management}}.
  \bibinfo{pages}{2486--2496}.
\newblock


\bibitem[Zhou et~al\mbox{.}(2022)]%
        {zhou2022simans}
\bibfield{author}{\bibinfo{person}{Kun Zhou}, \bibinfo{person}{Yeyun Gong},
  \bibinfo{person}{Xiao Liu}, \bibinfo{person}{Wayne~Xin Zhao},
  \bibinfo{person}{Yelong Shen}, \bibinfo{person}{Anlei Dong},
  \bibinfo{person}{Jingwen Lu}, \bibinfo{person}{Rangan Majumder},
  \bibinfo{person}{Ji-Rong Wen}, {and} \bibinfo{person}{Nan Duan}.}
  \bibinfo{year}{2022}\natexlab{}.
\newblock \showarticletitle{SimANS: Simple Ambiguous Negatives Sampling for
  Dense Text Retrieval}. In \bibinfo{booktitle}{\emph{Proceedings of the 2022
  Conference on Empirical Methods in Natural Language Processing: Industry
  Track}}. \bibinfo{pages}{548--559}.
\newblock


\bibitem[Zhuang et~al\mbox{.}(2021)]%
        {Zhuang2021DeepQL}
\bibfield{author}{\bibinfo{person}{Shengyao Zhuang}, \bibinfo{person}{Hang Li},
  {and} \bibinfo{person}{G. Zuccon}.} \bibinfo{year}{2021}\natexlab{}.
\newblock \showarticletitle{Deep Query Likelihood Model for Information
  Retrieval}. In \bibinfo{booktitle}{\emph{ECIR}}.
\newblock


\end{thebibliography}

\end{document}